%% file: main.tex
\documentclass[twoside]{article}

\usepackage{xcolor}
\usepackage{blindtext} % Package to generate dummy text throughout this template 
\usepackage{graphicx}
\usepackage[citebordercolor=white]{hyperref}
\usepackage{gensymb}
\usepackage[ruled,noend]{algorithm2e}
\usepackage{setspace}
\usepackage{framed,multirow}
\usepackage{amssymb}
\usepackage{cuted}
\usepackage{amsmath}
\usepackage{mathtools}
\usepackage[bb=dsserif]{mathalpha}
\usepackage{latexsym}
\usepackage{multicol}
\usepackage{microtype} % Slightly tweak font spacing for aesthetics

\usepackage[english]{babel} % Language hyphenation and typographical rules

\usepackage[hmarginratio=1:1,top=32mm,columnsep=20pt]{geometry} % Document margins
\usepackage[hang, small,labelfont=bf,up,textfont=it,up]{caption} % Custom captions under/above floats in tables or figures
\usepackage{booktabs} % Horizontal rules in tables

\usepackage{lettrine} % The lettrine is the first enlarged letter at the beginning of the text

\usepackage{enumitem} % Customized lists
\setlist[itemize]{noitemsep} % Make itemize lists more compact

\usepackage{abstract} % Allows abstract customization
\usepackage{titlesec} % Allows customization of titles
\titleformat{\section}[block]{\large\scshape\centering}{\thesection.}{1em}{} % Change the look of the section titles
\titleformat{\subsection}[block]{\large}{\thesubsection.}{1em}{} % Change the look of the section titles

\usepackage{fancyhdr} % Headers and footers
\usepackage{titling} % Customizing the title section
\usepackage{hyperref} % For hyperlinks in the PDF
\hypersetup{hidelinks = true}

\usepackage{expl3}
\ExplSyntaxOn
\tex_let:D \c_minus_one \scan_stop:
\int_const:Nn \c_minus_one {-1}
\ExplSyntaxOff

\usepackage{acro}
\acsetup{only-used=true}
\input{acro.tex}

\usepackage[round]{natbib}

\title{Automatic maneuver detection and tracking of space objects in optical survey scenarios based on stochastic hybrid systems formulation} % Article title
\author{%
\textsc{Guillermo Escribano}\thanks{Corresponding author. PhD candidate at Universidad Carlos III de Madrid}, \textsc{Manuel Sanjurjo-Rivo} \\[1ex] % Your name
\normalsize Universidad Carlos III de Madrid \\ % Your institution
\normalsize \href{mailto:guescrib@ing.uc3m.es}{guescrib@ing.uc3m.es}, \normalsize \href{mailto:msanjurj@ing.uc3m.es}{msanjurj@ing.uc3m.es} % 
\vspace*{0.5cm}\\%
\textsc{Jan Siminski} \\[1ex] % Your name
\normalsize ESA Space Debris Office \\ % Your institution
\normalsize \href{mailto:Jan.Siminski@esa.int}{Jan.Siminski@esa.int} %
\vspace*{0.5cm}\\%
\textsc{Alejandro Pastor, Diego Escobar} \\[1ex] % Your name
\normalsize GMV \\ % Your institution
\normalsize \href{mailto:apastor@gmv.com}{apastor@gmv.com}, \href{mailto:descobar@gmv.com}{descobar@gmv.com} %
}

\date{} % Leave empty to omit a date
\renewcommand{\maketitlehookd}{%
\begin{abstract}
\noindent 
\input{abstract}
\end{abstract}
}

\begin{document}
% Print the title
\maketitle

\vfill
\noindent
\textit{Keywords:} Space situational awareness, Maneuvering target tracking; Maneuver detection, Admissible region, Control distance metric, Stochastic hybrid systems
\vspace{0.5cm}
\newpage
%----------------------------------------------------------------------------------------
%	ARTICLE CONTENTS
%----------------------------------------------------------------------------------------

\input{sections/1_introduction}

\input{sections/2_problem_formulation}% stochastic hybrid systems
\input{sections/3_SHS_filtering} % filtering schemes for SHS

\input{sections/4_Maneuver_hypothesis_generation}% definition of admissible control region and plots
% brief description of two-impulse metric
% probably here we can include some notes on filtering, mainly referring to data association gates and stuff like that...
% definition of heurisitcs and how detected maneuvers can be used to approximate future ones... maybe talk about conditioned sampling using MCMC
\input{sections/5_results_and_comparison}

\input{sections/6_summary_and_conclusions}

\section*{Acknowledgements}

This work is part of an ongoing PhD thesis funded by the European Space Agency under the Networking Partnering Initiative through the Project \textit{Combined Heuristic and Statistical Methodologies applied to Maneuver Detection in the SST Observation Correlation Process} and also by the ``Comunidad de Madrid" under the Project %\textit{Métodos avanzados de correlación de medidas y determinación de órbita para la construcción y mantenimiento de un catálogo de objetos espaciales}.
\textit{Advanced measurement correlation and orbit determination methods for space object catalog build-up and maintenance}, grant number IND2017/TIC-7700.

%\nocite{*}
\bibliographystyle{plainnat}
\bibliography{refs}

\end{document}

%% file: acro.tex
\DeclareAcronym{DoD}{
	short = DoD,
	long  = United States Department of Defense,
	class = abbrev
}
\DeclareAcronym{DOT}{
	short = DOT,
	long  = United States Department of Transportation,
	class = abbrev
}
\DeclareAcronym{ITU}{
	short = ITU,
	long  = International Telecommunication Union,
	class = abbrev
}
\DeclareAcronym{NASDA}{
	short = NASDA,
	long  = National Space Development Agency of Japan,
	class = abbrev
}
\DeclareAcronym{COPUOS}{
	short = COPUOS,
	long  = Committee on the Peaceful Uses of Outer Space,
	class = abbrev
}
\DeclareAcronym{STSC}{
	short = STSC,
	long  = COPUOS Scientific and Technical Subcomittee,
	class = abbrev
}
\DeclareAcronym{IADC}{
	short = IADC,
	long  = Inter-Agency Space Debris Coordination Committee,
	class = abbrev
}
\DeclareAcronym{UNOOSA}{
	short = UNOOSA,
	long = Untied Nations Office for Outer Space Affairs,
	class = abbrev
}
\DeclareAcronym{USSTRATCOM}{
	short = USSTRATCOM,
	long  = United States Strategic Command,
	class = abbrev
}
\DeclareAcronym{JSpOC}{
	short = JSpOC,
	long  = Joint Space Operations Center,
	class = abbrev
}
\DeclareAcronym{JFCC SPACE}{
	short = JFCC SPACE,
	long = Joint Functional Component Command for Space,
	class = abbrev
}
\DeclareAcronym{NORAD}{
	short = NORAD,
	long = North American Aerospace Defense Command,
	class = abbrev
}
\DeclareAcronym{CNES}{
	short = CNES,
	long = Centre National d'Etudes Spatiales,
	class = abbrev
}
\DeclareAcronym{DLR}{
	short = DLR,
	long = German Aerospace Center,
	class = abbrev
}
\DeclareAcronym{ESA}{
	short = ESA,
	long = European Space Agency,
	class = abbrev
}
\DeclareAcronym{NASA}{
	short = NASA,
	long = National Aeronautics and Space Administration,
	class = abbrev
}
\DeclareAcronym{ROSCOSMOS}{
	short = ROSCOSMOS,
	long = State Space Corporation,
	class = abbrev
}
\DeclareAcronym{ComSpOC}{
	short = ComSpOC,
	long = Commercial Space Operations Center,
	class = abbrev
}
\DeclareAcronym{GTDS}{
	short = GTDS, 
	long = Goddard Trajectory Determination System,
	class = abbrev
}

\DeclareAcronym{SSN}{
	short = SSN,
	long  = Space Surveillance Network,
	class = abbrev
}
\DeclareAcronym{GRAVES}{
	short = GRAVES,
	long  = Grande Réseau Adapté à la Veille Spatial,
	class = abbrev
}
\DeclareAcronym{TIRA}{
	short = TIRA,
	long  = Tracking \& Imaging Radar,
	class = abbrev
}
\DeclareAcronym{ISON}{
	short = ISON,
	long  = International Scientific Optical Network,
	class = abbrev
}
\DeclareAcronym{S3T}{
	short = S3T,
	long  = Spanish Space Surveillance,
	class = abbrev
}
\DeclareAcronym{LEO}{
  short = LEO,
  long  = Low Earth Orbit,
  class = abbrev
}
\DeclareAcronym{MEO}{
  short = MEO,
  long  = Medium Earth Orbit,
  class = abbrev
}
\DeclareAcronym{GEO}{
  short = GEO,
  long  = Geostationary Earth Orbit,
  class = abbrev
}
\DeclareAcronym{GTO}{
	short = GTO,
	long  = Geostationary Transfer Orbit,
	class = abbrev
}
\DeclareAcronym{HEO}{
	short = HEO,
	long  = Highly Elliptical Orbit,
	class = abbrev
}
\DeclareAcronym{HAMR}{
	short = HAMR,
	long = High Area-to-Mass Ratio,
	class = abbrev
}
\DeclareAcronym{SRP}{
	short = SRP,
	long  = Solar Radiation Pressure,
	class = abbrev
}
\DeclareAcronym{NS}{
	short = NS,
	long  = North-South,
	class = abbrev
}
\DeclareAcronym{EW}{
	short = EW,
	long  = East-West,
	class = abbrev
}
\DeclareAcronym{ECI}{
	short = ECI,
	long  = Earth Centered Inertial reference frame,
	class = abbrev
}

\DeclareAcronym{COE}{
	short = COE,
	long  = Classical Orbital Elements,
	class = abbrev
}
\DeclareAcronym{OD}{
	short = OD,
	long  = Orbit Determination,
	class = abbrev
}
\DeclareAcronym{IOD}{
	short = IOD,
	long  = Initial Orbit Determination,
	class = abbrev
}
\DeclareAcronym{RMS}{
	short = RMS,
	long  = Root Mean Square,
	class = abbrev
}
\DeclareAcronym{WRMS}{
	short = WRMS,
	long  = Weighted Root Mean Square,
	class = abbrev
}
\DeclareAcronym{TLE}{
	short = TLE,
	long  = Two Line Element,
	class = abbrev
}
\DeclareAcronym{SGP}{
	short = SGP,
	long  = Simplified General Perturbations,
	class = abbrev
}
\DeclareAcronym{DSST}{
	short = DSST,
	long  = Draper Semi-analytical Satellite Theory,
	class = abbrev
}
\DeclareAcronym{AMR}{
	short = AMR,
	long  = Area-to-Mass Ratio,
	class = abbrev
}
\DeclareAcronym{MTT}{
	short = MTT,
	long = Multi Target Tracking,
	class = abbrev
}
\DeclareAcronym{JPDA}{
	short = JPDA,
	long = Joint Probabilistic Data Association,
	class = abbrev
}
\DeclareAcronym{MHF}{
	short = MHF,
	long = Multiple Hypothesis Filter,
	class = abbrev
}
\DeclareAcronym{MHT}{
	short = MHT,
	long = Multiple Hypothesis Tracking,
	class = abbrev
}
\DeclareAcronym{RFS}{
	short = RFS,
	long = Random Finite Set,
	class = abbrev
}
\DeclareAcronym{FISST}{
	short = FISST,
	long = Finite Set Statistics,
	class = abbrev
}
\DeclareAcronym{EKF}{
	short = EKF,
	long = Extended Kalman Filter,
	class = abbrev
}
\DeclareAcronym{UKF}{
	short = UKF,
	long = Unscented Kalman Filter,
	class = abbrev
}
\DeclareAcronym{CAR}{
    short = CAR,
    long = Constrained Admissible Regions,
    class = abbrev
}
\DeclareAcronym{DA}{
    short = DA,
    long = Differential Algebra,
    class = abbrev
}
\DeclareAcronym{BPF}{
    short = BPF,
    long = Bootstrap Particle Filter,
    class = abbrev
}
\DeclareAcronym{KDE}{
    short = KDE,
    long = Kernel Density Estimator,
    class = abbrev
}
\DeclareAcronym{MH}{
    short = MH,
    long = Metropolis-Hastings,
    class = abbrev
}
\DeclareAcronym{ML}{
    short = ML,
    long = Machine Learning,
    class = abbrev
}
\DeclareAcronym{BLS}{
    short = BLS,
    long = Batch Least-Squares,
    class = abbrev
}
\DeclareAcronym{DREAM}{
    short = DREAM,
    long = DiffeRential Evolution Adaptive Metropolis,
    class = abbrev
}
\DeclareAcronym{SK}{
    short = SK,
    long = Station Keeping,
    class = abbrev
}
\DeclareAcronym{MEE}{
    short = MEE,
    long = Modified Equinoctial Elements,
    class = abbrev
}
\DeclareAcronym{MLE}{
    short = MLE,
    long = Maximum Likelihood Estimate,
    class = abbrev
}
\DeclareAcronym{SHS}{
    short = SHS,
    long = Stochastic Hybrid System,
    class = abbrev
}
\DeclareAcronym{RMSE}{
    short = RMSE,
    long = Root Mean Square Error,
    class = abbrev
}
\DeclareAcronym{NEES}{
    short = NEES,
    long = Normalized Estimation Error Squared,
    class = abbrev
}
\DeclareAcronym{PCRB}{
    short = PCRB,
    long = Posterior Cramér Rao Bound,
    class = abbrev
}
\DeclareAcronym{IMM}{
    short = IMM,
    long = Interacting Multiple Models,
    class = abbrev
}
\DeclareAcronym{GMM}{
    short = GMM,
    long = Gaussian Mixture Model,
    class = abbrev
}
\DeclareAcronym{PoL}{
    short = PoL,
    long = Patterns of Life,
    class = abbrev
}
\DeclareAcronym{T2T}{
    short = T2T,
    long = track-to-track,
    class = abbrev
}
\DeclareAcronym{T2O}{
    short = T2O,
    long = track-to-orbit,
    class = abbrev
}
\DeclareAcronym{O2O}{
    short = O2O,
    long = orbit-to-orbit,
    class = abbrev
}
\DeclareAcronym{SMC}{
short = SMC,
long = Sequential Monte Carlo,
class = abbrev
}
\DeclareAcronym{PF}{
    short = PF,
    long = Particle Filter,
    class = abbrev
}
\DeclareAcronym{MCMC}{
    short = MCMC,
    long = Markov Chain Monte Carlo,
    class = abbrev
}
\DeclareAcronym{NLP}{
    short = NLP,
    long = Non-Linear Programming,
    class = abbrev
}
\DeclareAcronym{SST}{
	short = SST,
	long  = Space Surveillance and Tracking,
	class = abbrev
}
\DeclareAcronym{SSA}{
	short = SSA,
	long  = Space Situational Awareness,
	class = abbrev
}
\DeclareAcronym{FOV}{
	short = FOV,
	long  = Field of View,
	class = abbrev
}
\DeclareAcronym{FOR}{
	short = FOR,
	long  = Field of Regard,
	class = abbrev
}
\DeclareAcronym{STM}{
	short = STM,
	long  = Space Traffic Management,
	class = abbrev
}
\DeclareAcronym{OTOA}{
	short = OTOA,
	long  = Observation-To-Observation Association,
	class = abbrev
}
\DeclareAcronym{TTTA}{
	short = TTTA,
	long  = Track-To-Track Association,
	class = abbrev
}
\DeclareAcronym{OTTA}{
	short = OTOA,
	long  = Observation-To-Track Association,
	class = abbrev
}
\DeclareAcronym{SATCAT}{
	short = SATCAT,
	long  = Satellite Catalogue,
	class = abbrev
}
\DeclareAcronym{RSO}{
	short = RSO,
	long = Resident Space Object,
	class = abbrev
}
\DeclareAcronym{TSA}{
	short = TSA,
	long = Too Short Arc,
	class = abbrev
}
\DeclareAcronym{VSA}{
	short = VSA,
	long = Very Short Arc,
	class = abbrev
}
\DeclareAcronym{UCT}{
	short = UCT,
	long = uncorrelated track,
	class = abbrev
}
\DeclareAcronym{UCO}{
	short = UCO,
	long = uncorrelated optical observation,
	class = abbrev
}
\DeclareAcronym{SIOD}{
	short = SIOD,
	long = Statistical Initial Orbit Determination,
	class = abbrev
}

\DeclareAcronym{PDF}{
	short = PDF,
	long = Probability Distribution Function,
	class = abbrev
}
\DeclareAcronym{CDF}{
	short = CDF,
	long = Cummulative Distribution Function,
	class = abbrev
}

%% file: abstract.tex
%%%
%Maneuver detection of Resident Space Objects is an active research topic, highly demanded in Space Surillance and Tracking applications. The present work proposes a novel approach for tracking maneuverable objects in a reduced observability scenario. The latter is typically the case for Medium or Geostationary Earth Orbits, altitudes at which only optical observations are usually available. The novelty of the method relies on the use of a Sequential Monte Carlo framework for state estimation, enhanced by Markov Chain Monte Carlo post-maneuver state recovery. Through the definition of surrogate functions to approximate maneuvers, it is possible to recover a region of states compliant with the incoming observations provided some bounds for the maximum expected control effort. Additionally, hypotheses can be formulated based on historical data or otherwise already characterized maneuvers. Results are obtained for Geostationary spacecraft in a simulated optical observation scenario.
%%%%
The state space representation of active resident space objects can be posed in the form of a stochastic hybrid system. Satellite maneuvers may be accounted for according to control cost or heuristical considerations, yet it is possible to jointly consider a combination of both. In this work, Sequential Monte Carlo filtering techniques are applied to the maneuvering target tracking problem in an optical survey scenario, where the maneuver control inputs are characterized in a Bayesian inference process. Due to the scarcity of data inherent to space surveillance and tracking, model switching probabilities are not estimated but derived from the ability of the state representation to fit incoming measurements. A Markov Chain Monte Carlo sampling scheme is used to explore the region assumed accessible to the object in terms of the hypothesized post-maneuver observation and a novel and efficient control distance metric. Results are obtained for a simulated optical survey scenario, and comparisons are drawn with respect to a moving horizon least-squares estimator. The proposed framework %shows promising results towards automating online maneuver detection and data association
is proved to allow for a capable implementation of an automated online maneuver detection algorithm, thus contributing to the reduction of uncertainty in the state of active space objects. % and a multi-hypothesis filter based on a bank of Kalman filters.
% The abstract is identified by the word Abstract. A concise and factual abstract is required. The abstract should state briefly the purpose of the research, the principal results and major conclusions. Abstracts are extracted for abstracting services, and each abstract should summarize the content of the article. 

%% file: sections/1_introduction.tex
\section{Introduction}\label{sec:intro}

The interest in \ac{SST} has been steadily growing in the last two decades, as the effects of Earth orbital congestion become more evident. Means to characterize the Earth's orbital population are then required in order to ensure a safe and orderly growth of space activities. Accordingly, sensor data retrieved by surveillance radars and optical telescopes is used to build up and maintain catalogs of Earth orbiting objects. \ac{SST} cataloging systems are in charge of processing incoming measurements, thereby updating the state of cataloged objects or detecting new ones. These measurements are usually obtained as a time series of observations that stem from a common object, also known as \textit{tracks}. If the distance between a well-established orbit of a known object and a given track is sufficiently small, then such set of observations is assigned (\textit{correlated}) to the object and its corresponding orbital state is updated. On the contrary, the track is categorized as \textit{uncorrelated} and compared against other uncorrelated tracks. %The latter process involves evaluating 
Different combinations of uncorrelated tracks are evaluated until a sufficiently high number (typically $\sim 3$-$4$, see \cite{hill2012comparison}) is compliant with a specific orbit, resulting in a new object identification. % i.e. a new object can be identified.
%determining whether they belong to an already cataloged object, associate with previous uncorrelated tracks (potentially corresponding to a new object detection) or cannot be simply correlated.
  Operational or active spacecraft perform frequent maneuvers, usually dictated by mission requirements or collision avoidance, potentially hindering the data association problem. In the absence of capable maneuver detection and estimation methods, post-maneuver measurements may trigger the identification of a new (duplicated) object as part of the catalog build-up and maintenance process. To overcome this limitation, the space accessible to an active spacecraft can be expressed in terms of a control effort (\cite{holzinger2011optimal}), with a post-maneuver state probability distribution based on the control effort itself or otherwise previously characterized maneuvers. Uncertainty regarding the state of active spacecraft poses a threat to data association and tracking of space object catalogs. Incorporating knowledge regarding the maneuver history of this active population can aid in providing better predictions for their state, partially contributing to the reduction of epistemic uncertainty at a \ac{STM} system level (see \cite{DELANDE20181800}, \cite{uncertaintySSA}).

Various efforts have been directed in this regard, initially to identify maneuvers performed by known objects. \cite{kelecy2007satellite} devised a method to detect maneuvers based on changes in \ac{TLE} data: by comparing orbits reported at different epochs one could identify \textit{non-environmental} perturbations, i.e. thrust arcs. The interest in maneuver detection then moved towards solving the data association problem. The former work by \cite{Holzinger2012}, but also \cite{Lubey}, reduces the combined maneuver detection and data association problem to solving the optimal control input required to be compliant with a certain observation given a prior orbit. The aim is not to characterize the maneuver in terms of control input, but to provide a control distance metric that can be used for hypothesis testing, i.e. decide whether an incoming observation is triggered by a maneuvered object. A thorough implementation of this approach, with certain modifications, has been recently explored by \cite{SERRA2021}, whose main focus is on the data association problem at a \ac{SST} system level considering multiple targets. Additional efforts have been directed towards jointly solving the data association and maneuver estimation, as proposed by \cite{pastormaneuver}. Therein, maneuvers are estimated in the form of impulsive burns performed within a discrete temporal grid, where the aim is to minimize a cost function weighting the distance to the observations and the required control effort in terms of $\Delta V$ magnitude. In parallel, an increasing interest was given to the so-called \textit{patterns of life} (cf. \cite{Cox2016TheSO}). In general, spacecraft maneuver to maintain their location within certain orbital slots, so the use of previously characterized maneuvers may result in an increased performance in terms of maneuver detection and state estimation. \ac{ML} techniques are well suited to this type of applications, and implementations can be found in the works of \cite{shabarekhML}, \cite{Singh2016Athena} and \cite{abay2018maneuver}. Not only \ac{ML} techniques have been applied to this end, but also former statistical methodologies. \cite{siminski2017correlation} propose the use of a \ac{KDE} to characterize maneuvers in terms of the relative change in orbital elements and the pre-maneuver orbital state. In this approach, data association is based on the information contained within the \ac{KDE} and a control distance metric $\Delta V$. An admissible region is defined based on a maximum threshold $\Delta V_{max}$ so the generated post-maneuver state hypotheses are tested to lie within such region.

\begin{figure*}[!htb]
    \centering
    \includegraphics[trim = 0 30 0 0, clip, width = \linewidth]{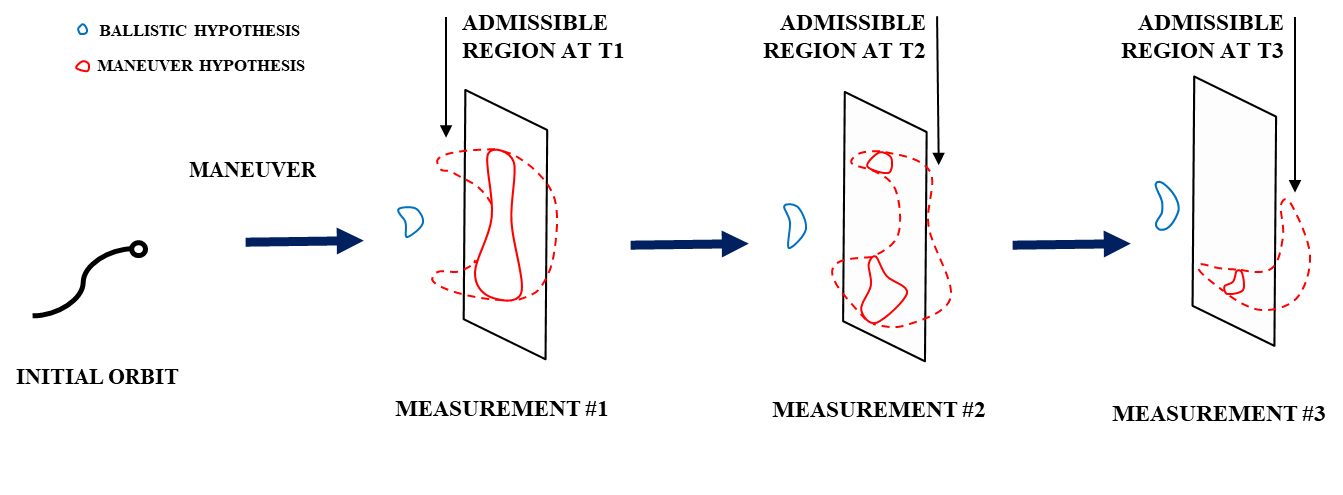}
    \caption{Sketch of the measurement association sequence in the event of a maneuver. The space accessible to the object is given by the admissible control region.}
    \label{fig:assocSequence}
\end{figure*}

% 1) stochastic hybrid systems
In this paper, the authors propose the use of a \ac{SHS} (cf. \cite{lygeros2000}) formulation as it conforms a natural statistical framework for the modelling of systems for which there are multiple accessible dynamical models, as is the case for maneuvering targets. The flexibility inherent to this formulation allows fusing information coming from very different sources, e.g. knowledge derived from historical data and optimal control.
% 2) filtering methods for SHS: sequential monte carlo and MCMC
State space filtering of \ac{SHS} is usually solved by means of multiple model gaussian (cf. \cite{IMMCKF}, \cite{IMM2}) and non-gaussian (cf. \cite{EPFJMS}) algorithms. Due to the non-linearity of the space environment and the special characteristics of optical survey scenarios %(\cite{tflohrerOpticalSurvey})
(\cite{PFSOT}), a \ac{SMC} scheme is used to estimate or infer the state and active mode of the system conditioned on a sequence of observations. Our proposal takes the form of a bi-modal system, in which either the non-maneuvering or maneuvering mode is active prior to the last hypothesized observation. In the non-maneuver mode, deterministic dynamical models usually provide sufficiently accurate approximations. However, in the presence of maneuvers, the control input to the system is indirectly inferred through its effects on the state. To this end, \ac{MCMC} techniques are used to explore the posterior state distribution conditioned on different prior assumptions, effectively conforming alternative hypotheses. These hypotheses are then pruned, promoted and merged, ultimately converging to the true association sequence as more information is available. This procedure is illustrated in Fig. \ref{fig:assocSequence}, which depicts the post-maneuver measurement association sequence in the state space, clearly distinguishing between the ballistic and maneuvered accessible regions.
% 3) "hypothesis" generation: definition of admissible control region and control distance metric
The accessible space in the event of a maneuver, termed \textit{admissible control region}, defines the boundaries for \ac{MCMC} sampling. The latter can be regarded as 
%Within this paper, the authors propose a combined methodology to fuse information coming from both sources, i.e. the control cost and the maneuver history. To this end,
a revision of the \textit{admissible region} (cf. \cite{milaniAR}) based on the control effort required to transfer from an initial to a final orbit.
The requirements imposed by the definition of the admissible control region and \ac{MCMC} sampling lead to the derivation of a novel and efficient control distance metric, $P$, which allows for a fast computation of the control cost separating two orbits within a specific time of flight.
% 4) generation and use of heuristics and how they can improve state estimation and data association performance
Under the assumption of optimal maneuvers in a control cost sense, one can elaborate a candidate post-maneuver state distribution function with higher probability densities in lower $P$ regions. However, note that the orbital information contained within a few optical observations is limited, and so the optimal transfer assumption may not be sufficient to provide an estimate of the post-maneuver orbital state. The use of prior knowledge, i.e patterns of life, can improve the estimation performance given the target follows a repetitive maneuvering plan.
Contributions of this work are thus not limited to a combined optimal-heuristic maneuver detection but also a realistic uncertainty characterization of the state of active space objects following a Bayesian inference process.
%This is enabled by 1) the definition of an efficient control distance metric $P$ that allows for a fast computation of the control cost separating two orbits and 2) a recursive generation of heuristics describing the maneuvering characteristics and orbital boundaries of an object.

The content of the paper is structured as follows. The problem and its mathematical formulation is discussed in Section \ref{sec:prob}. %ac{SMC} methods are introduced in Section \ref{sec:smc}, wherein a brief description of state space filtering can also be found. Section \ref{sec:MCMC} is devoted to \ac{MCMC} sampling techniques, emphasizing on their use as state estimation methods in the absence of properly characterized dynamical models.
Different methods for estimating the state of stochastic hybrid systems are discussed in Section \ref{sec:fshs}. Special emphasis is given to non-linear non-gaussian filtering both in the presence and absence of a proper definition for the underlying dynamical model.
%The admissible control region concept is presented in Section \ref{sec:arc}, in analogy to the former definition of the constrained admissible region. Means to numerically approximate this admissible set are also provided, and the definition of an efficient control distance metric that can be used to derive such region is briefly discussed in Section \ref{sec:control}. Section \ref{sec:heuristics} deals with the treatment and generation of maneuver heuristics, which are largely based on the relative change in orbital parameters caused by the maneuver.
Section \ref{sec:mhg} deals with the generation of maneuver hypotheses, i.e. post-maneuver state estimation based on alternative prior assumptions. In this regard, the space accessible to the system is bounded in terms of a maximum expected control effort, derived by means of a novel control distance metric. Hypotheses are then generated via \ac{MCMC} sampling based on optimal control or conditioned on the patterns of life of the target.
Results for simulated optical observations are presented and analyzed in Section \ref{sec:results}, where the filter performance is compared against different maneuvering target tracking filter implementations. Finally, conclusions are drawn in Section \ref{sec:conclusions} together with guidelines for future lines of research.

%% file: sections/2_problem_formulation.tex
\section{Problem formulation}\label{sec:prob}

The problem of maneuver detection and data association can be regarded as a maneuvering target tracking problem if required to be solved in an automated manner, whose mathematical description can be naturally framed under the stochastic hybrid systems formulation.
%and  The stochastic hybrid systems formulation provides a natural framework for the mathematical description of the state evolution of a maneuvering target.
Herebelow, the general description of a \ac{SHS} is first presented, and then particularized to the case of maneuvering space object tracking based on optical observations.

\subsection{Stochastic hybrid systems}

The general stochastic hybrid state space filtering problem is governed by the following system of stochastic difference equations (cf. \cite{blomBloemSHS}; \cite{lygerosSHS}):
\begin{align}
    & \label{eq:statediff} \mathbf{x}_k = f(\mathbf{x}_{k-1},\mathbf{u}_k,m_{k},t_{k})+g(\mathbf{x}_{k-1},\mathbf{u}_k,m_{k},t_{k})\mu_k \, , \\
    & m_{k} = \pi(\mathbf{x}_{k-1},\mathbf{u}_k,m_{k-1},t_k,\nu_k) \,,\\
    & \mathbf{z}_k = h(\mathbf{x}_{k},t_{k})+q(\mathbf{x}_{k},t_{k})\gamma_k \,,\label{eq:zofx}
\end{align}
where $\mathbf{x}_k \in \mathbb{R}^n$ represents the physical state of the system, or continuous valued variables being estimated, $\mathbf{u}_k$ is the control input sequence from $t_{k-1}$ to $t_k$, and $m_k \in \mathbb{M}$ refers to the discrete valued system mode active at time $t_k$. The mapping function $f(\cdot)$ represents the underlying dynamical model, which typically follows a system of first order differential equations. $g(\cdot)$, usually referred to as diffusion coefficient, is included in order to account for any kind of colored noise for the Wiener process $\mu_k$, and is commonly approximated in a scalar or matrix form. Mode transitions are given by the mapping $\pi(\cdot)$, where $\nu_k$ is an independent random variable representing the stochastic process, analogous to $\mu_k$ in Eq. \eqref{eq:statediff}. The observations available to the system are indicated by $\mathbf{z}_k \in \mathbb{R}^p$, in which $h(\cdot)$ is the deterministic non-linear mapping between the system internal and observed states, and $q(\cdot)$ is again introduced to allow for any type of noise model realized through the random variable $\gamma_k$.

\subsection{Maneuvering space object tracking}
\label{sec:msotrack}

For the intended application, the state of the system and observed quantities read
\begin{equation}
    \mathbf{x}_{k} = \begin{pmatrix} \mathbf{r}_k \\ \dot{\mathbf{r}}_k \\ B_k \end{pmatrix} \ \ \ \ \ \text{and} \ \ \ \ \  \mathbf{z}_{k} = \begin{pmatrix} \alpha_k \\ \delta_k \\ \dot{\alpha}_k \\ \dot{\delta}_k \end{pmatrix},
\end{equation}
being $\mathbf{r}_k$ the Cartesian representation of the position vector, $\mathbf{\dot{r}}_k$ its temporal derivative at time $t_k$ and $B_k$ the solar radiation pressure coefficient of the target object. The latter is included in order to account for variations in the area to mass ratio (exposed to the sun) that stem from changes in attitude (orientation) and mass expenditure due to maneuvers, among others. Note the focus is on optical observations, which are mainly used to track high altitude objects. Thereafter, no atmospheric effects are considered as the atmospheric density is assumed to have a negligible impact on the system dynamics.

Observation data from optical sensors is retrieved as a time series, i.e. track, of correlated right ascension $\alpha$ and declination $\delta$ pairs, which are expressed in a topocentric reference frame centered at the sensor location. These correlated pairs conform the so-called optical tracks, whose typical duration spans from 2 to 10 minutes depending on the survey strategy. The orbital period of the observed objects is usually of the order of days or tens of hours, so the orbital arc described within a track is relatively small ($<1$\%). %, and have a typical duration of 2 to 10 minutes depending on the survey strategy.
Information contained within such short tracks is therefore limited so they are usually approximated by a linear regression at the mean epoch. The output of this linear regression is commonly referred to as \textit{attributable}, and was first proposed in \cite{milaniAR}. An attributable $(\alpha,\delta,\dot{\alpha},\dot{\delta})$ can then be used to define the line-of-sight of an object
\begin{equation}
    \mathbf{w}_k = \begin{pmatrix} \cos{\alpha_k} \cos{\delta_k} \\ \sin{\alpha_k} \cos{\delta_k} \\
    \sin{\delta_k} \end{pmatrix}
\end{equation}
%is the line-of-sight of the object as seen from a ground-based telescope, being $\alpha$ the right ascension and $\delta$ the declination in a topocentric reference frame centered at the sensor location. The temporal derivative of the line-of-sight
and its temporal derivative
\begin{equation}
    \dot{\mathbf{w}}_k = \dot{\alpha}\begin{pmatrix} -\sin{\alpha_k} \cos{\delta_k} \\ \cos{\alpha_k} \cos{\delta_k} \\
    0 \end{pmatrix}+\dot{\delta}\begin{pmatrix} -\cos{\alpha_k} \sin{\delta_k} \\ -\sin{\alpha_k} \sin{\delta_k} \\
    \cos{\delta} \end{pmatrix}.
\end{equation}
%is solely defined in terms of the angles and angle rates.
%Note the adoption of the \textit{attributable} format for the observations (see \cite{milaniAR}), which is the result of a linear least-squares fit on a series of correlated $\alpha$-$\delta$ pairs.
The line-of-sight and its temporal derivative are related to the state variables via
\begin{align} \label{eq:rho}
    & \mathbf{r} = \mathbf{r_s} + \rho \mathbf{w} \\
    & \mathbf{\dot{r}} = \mathbf{\dot{r}_s} + \rho \mathbf{\dot{w}} + \dot{\rho}\mathbf{w} \label{eq:drho},
\end{align}
where $\mathbf{r_s}$ and $\mathbf{\dot{r}_s}$ are the position and velocity of the observing site, $\rho$ is the range or positional distance between the object and the sensor, and $\dot{\rho}$ its time derivative.

With regard to the system dynamics, the mapping $f(\cdot)$ in Eq. \eqref{eq:statediff} is generally approximated by a dynamical model that takes the form of a perturbed restricted two body problem, with the Earth as central body. Therein, the system is assumed to be a point particle upon which different forces and perturbations act. Besides the central gravity field exerted by the Earth, the effect of Sun, Moon and planets are also considered, together with a non-spherical Earth model and solar radiation pressure: recall atmospheric effects are disregarded due to the orbital regions of interest in optical surveys. Approximating reality in a purely deterministic fashion requires an extensive modelling effort, including but not limited to the roto-translational effects (c.f. \cite{fruh2013coupled}, \cite{COAESnearEarth}). To partially mitigate the error introduced by mismodeled dynamics, it is common practice to introduce process noise, which in this case is represented by the second term in the right-hand side of Eq. \eqref{eq:statediff}.
%, usually referred to as process noise, is introduced to account for mismodeled dynamics.

The subset $\mathbf{u}_k$ represents the control input to the system and is related to the active mode $m_k$ by
\begin{equation}\label{eq:mode}
m_k \equiv \begin{cases} \mbox{$\emptyset$,} & \mbox{if $\mathbf{u}_k = 0 $} \\ \mbox{$\mathbb{1}$,}& \mbox{otherwise} \end{cases}
\end{equation}
Due to the scarcity of data inherent to optical survey scenarios, it is convenient to estimate the state after the maneuver $\mathbf{x}_k$ instead of the control sequence $\mathbf{u}_k$, which involves inferring the active mode $m_k$ at time $t_k$. Accordingly, the proposed filtering problem considers maneuver detection and tracking but not the estimation of the control input to the system.

%replace the estimation of the control sequence $\mathbf{u}^*$ by its contribution to the remaining state variables $( \mathbf{r}, \dot{\mathbf{r}}, \Gamma )$. 

%For the problem of maneuvering space object tracking, the maneuver set $\mathbb{M}$ can be characterized by the finite maneuvering modes that are accessible to the system. An example would be a typical \ac{GEO} mission, in which the satellite performs east-west (EW) and north-south (NS) station keeping maneuvers to remain within its intended operational latitude and longitude slot (see e.g. \cite{stationKeeping}). In this example, the cardinality of $\mathbb{M}$, denoted by $|\mathbb{M}|$, would consider the empty set (no maneuver scenario) and all the possible realizations of station keeping maneuvers, i.e. $m_k\in \left\{ \emptyset , EW^+, EW^-, NS^+, NS^- \right\}$.

%% file: sections/3_SHS_filtering.tex
\section{State estimation of stochastic hybrid systems}\label{sec:fshs}

% Introduce the content of the section:
% Here we propose a means to perform state space filtering of the SHS described in section 2. Hereunder we briefly introduce the concept of state space filtering and different filtering techniques that can be applied to non-linear systems. We also need to mention how we handle the duality of the system, i.e. in the non-maneuvering mode we follow a SMC with the system dynamics and in the maneuver mode we generate samples from a post-maneuver state distribution conditioned on 1) the control distance from the post-maneuver state to the pre-maneuver orbit and 2) heuristics derived from the maneuver history. The latter represent different maneuver hypothesis whose sampled distributions are determined from the application of an advanced MCMC technique. 

Within this paper we propose a method to perform state space filtering for the \ac{SHS} described in Section \ref{sec:prob}. Hereunder, we briefly introduce the concept of state space filtering and discuss on different filtering techniques commonly applied to state estimation of non-linear systems. Bi-modality of the system is considered at an upper filtering level, resulting in different schemes for estimating the state in the non-maneuvering and maneuvering modes. In the former case we propose a \ac{SMC} method following the \textit{known} or \textit{ballistic} system dynamics. In the maneuver mode we generate samples from two different post-maneuver state distributions, conditioned on 1) the control distance from the post-maneuver state to the pre-maneuver orbit and 2) heuristics derived from the maneuver history, as detailed in Section \ref{sec:mhg}. The latter procedure results in different maneuver hypotheses whose sampled distributions are determined from the application of an advanced \ac{MCMC} technique, i.e. the \ac{DREAM} algorithm. 

\subsection{State space filtering}
State space filtering refers to the process of estimating the optimal sequence of states $\mathbf{x}_{0:N_z}$ according to a set of system observations $\mathbf{z}_{1:N_z} = \mathbf{z}(\mathcal{T})$ at times $\mathcal{T}= \{t_1,t_2,...,t_{N_z}\}$. Numerous techniques can be applied to the aforementioned problem attending to the mathematical behavior of the system of interest. Under linear dynamics and Gaussian unbiased noise assumptions, an analytical solution exists, as is the well-known Kalman Filter. This scheme has been successfully applied to non-linear systems under appropriate transformations, see the work in \cite{UKF} and \cite{CKF}. Based on the assumption of Markovian dynamics, the Kalman filter adopts a sequential scheme so that the estimation is assumed to be optimal at each observation and the sequence of state estimates is not refined as new observations arrive. Another class of estimation methods, termed \ac{BLS}, are aimed at minimizing the squared distance to the observations. Under the assumption of a Gaussian state probability distribution, the estimation is performed on a batch of measurements. These methods do not target to solve the state estimation problem in a sequential manner but are rather aimed at fitting a dynamical model to a set of observations. Note however,  under certain modifications they can be readily applied to sequential estimation, as suggested in \cite{BLS} for the moving horizon estimator.
%Batch Least-Squares (see Section \ref{NLBLS}) methods, on the contrary, are aimed at fitting a finite set of observations to the model in a least squares sense, yet under the assumption of Gaussian probability distribution of the state.

Whenever the probability distribution function (pdf) of the state is not known \textit{a priori} (or it is subject to non-linear transformations), the complexity of the estimation problem significantly increases. This is to say, if the Gaussian assumption $\mathbf{x}(t)\sim p(\mathbf{x}(t),t) \approx \mathcal{N}\begin{pmatrix}\mathbf{x}(t); \hat{\mathbf{x}}(t),\Sigma(t)\end{pmatrix}$, where $\hat{\mathbf{x}}$ and $\Sigma$ are the expected state estimate and co-variance, is dropped, then the evolution of the probability distribution function $p(\mathbf{x}(t),t)$ with time is governed by the Fokker-Planck-Kolmogorov equation, Eq. \eqref{eq:FPKEquation}, which makes reference to the work by \cite{fokker} and \cite{planck}.
\begin{equation}\label{eq:FPKEquation}
\begin{array}{ll}
\displaystyle
\frac{\partial p(\mathbf{x}(t),t)}{\partial t} = & -\displaystyle \sum_{i=1}^{n} \frac{\partial}{\partial x_i}\begin{bmatrix}
f(\mathbf{x}(t),t)p(\mathbf{x}(t),t)
\end{bmatrix}
\\ & \displaystyle+\sum_{i=1}^{n}\sum_{i=1}^{n}\frac{\partial^2}{\partial x_i\partial x_j}\begin{bmatrix}
w(t)p(\mathbf{x}(t),t) . 
\end{bmatrix}

\end{array}
\end{equation}
The term $f(\cdot)$ in Eq. \eqref{eq:FPKEquation} follows the definition given in Eq. \eqref{eq:statediff}, whereas $w(t)$ corresponds to the second term of the right hand side of Eq. \eqref{eq:statediff}.

This complicated partial differential equation is often approximated using Monte Carlo integration (cf. \cite{montecarlomethod}), e.g. via \ac{SMC} or \ac{MCMC} techniques. Disregarding the assumption of a Gaussian state distribution leads to a re-definition of how the measurement information is introduced into the system. Kalman and Batch Least-Squares filters use the observed quantities to reduce the state co-variance and update the expected value (or state estimate). However, if the state distribution is allowed to take any realization, the update needs to be performed on the entire distribution. 

\subsection{Sequential Monte Carlo}\label{sec:smc}

SMC methods are referred to as \acp{PF} when applied to filtering problems. The term \textit{particle} stems from their approach to managing non-linear transformations of statistical distributions, i.e. approximating Eq. \eqref{eq:FPKEquation}. Given an initial distribution, a statistically significant number of samples $N$ are randomly drawn, approximating the initial pdf as a sampled distribution
\begin{equation}
    p(\mathbf{x}) \approx p_s(\mathbf{x}) = \sum_{i=1}^N\omega_i \delta(\mathbf{x}-\mathbf{x}_i),
\end{equation}
where $\omega_i$ is the weight associated to sample $\mathbf{x}_i$ and $\delta(\cdot)$ is the Dirac delta function.
 Each of these samples, or particles, is then propagated following the system dynamics so the non-linear transformation can be fully characterized at least in the statistical region of interest. This procedure results in an approximation of the \textit{a posteriori} distribution of the state that converges to the true solution as the number of particles is increased, i.e. $N\rightarrow\infty$. Conceptually, \acp{PF} are appealing since at a higher computational cost, it is possible to perform an arbitrarily accurate non-linear propagation of the state uncertainty. 

\subsubsection{Bootstrap particle filter}\label{sec:bpf}

The combination of Sequential Importance Sampling and Resampling (SISR) conforms the \ac{BPF}, which is in fact the simplest realization of a capable SMC method for state estimation. Sequential importance sampling refers to the process of updating the sampled state distribution according to the information conveyed in the observation sequence. This update is reflected on the weight associated to each particle, so that at some point, a given particle may have negligible weight. The latter problem can be mitigated through the use of resampling. Sampling from an already sampled distribution results in a population with a lower weight variability, eliminating particles with smaller weights and duplicating those with higher probability.

%In fact, there are applications, such as space surveillance, in which the dynamical model $f(\cdot)$ can be relatively expensive to evaluate, a priori hindering the applicability of this type of filters. It is, however, important to recall that each particle can be propagated independently so the use of parallel and distributed computing is widely adopted. In fact, opposed to methods based on optimization or differential algebra, the computational time of a particle filter can \textit{a priori} be estimated so an on-line PF implementation is feasible given the computational burden can be afforded.

Algorithm \ref{algo:BPF} summarizes the main steps required by the \ac{BPF} scheme, which is particularly simple and intuitive from a statistical perspective. These type of filters usually require a significant number of particles $N_{MC}$ to operate and, as such, they suffer from the \textit{curse of dimensionality} (see \cite{curseofdimensionality}): the sample size is required to grow exponentially with the number of state variables to avoid collapse, which is expected to occur as $log(N_{MC})/n\rightarrow 0$. Applications to high-dimensional systems also feature faster particle depletion rates, so naive resampling techniques as the one in Algorithm \ref{algo:BPF} may lead to a population that is concentrated on a single particle for relatively low dynamical noise levels.

\begin{algorithm}[h]
\SetAlgoLined
\setstretch{1.2}
%\KwResult{Write here the result
\textsc{Initialization:} sample $N$ particles $\mathbf{x}_{0,i}$ from $p(\mathbf{x}_0)$ with weights $\omega_i = \frac{1}{N}$\\
\For{$k>1$}{
$\ \ $\textsc{Importance Sampling}\\
1) Approximate $p(\mathbf{x}_k|\mathbf{z}_{1:k-1})$: \\
$\ \ \ \ \ \omega_{i,k}^-=\omega_{i,k-1} \ \ \ \ \ \mathbf{x}_{k,i}^- = f(\mathbf{x}_{k-1,i},w_i,t_{k-1})$\\ \vspace*{0.1cm}
2) Apply the measurement update to obtain $p(\mathbf{x}_k|\mathbf{z}_{1:k})$:\\ \vspace*{0.2cm}
$\ \ \ \ \ \omega_{i,k} =\displaystyle  \frac{\omega_{i,k}^-p(\mathbf{z}_k|\mathbf{x}_{k,i})}{\sum_{i=1}^N\omega_{i,k}^-p(\mathbf{z}_k|\mathbf{x}_{k,i})} \ \ \ \ \ \mathbf{x}_{k,i} =\mathbf{x}_{k,i}^-$\\ \vspace*{0.2cm}
$\ \ $\textsc{Resampling}\\ \vspace*{0.1cm}
3) Compute the Effective Sample Size: \\
$\ \ \ \ \ ESS = \displaystyle \frac{1}{\sum_{i=1}^N\omega_{i,k}^2}$\\
\If{$ESS\leq ESS_{min}$}{ \vspace*{0.1cm}
4) Compute the cumulative distribution of $p_s(\mathbf{x}_k|\mathbf{z}_{1:k})$:\\
$\ \ \ \ \ P_s(\mathbf{x}_k|\mathbf{z}_{1:k}) = \displaystyle\sum_{i\in\mathcal{I}(x)}\omega_{i,k} \ \ \ \ \ \mathcal{I}(x) = \{i:x_k \leq x\}$\\ \vspace*{0.1cm}
5) Draw $u_i$ from $\mathcal{U}(0,1)$ and update the particles according to:\\
$\ \ \ \ \ \omega_{i,k} = \frac{1}{N} \ \ \ \ \ \mathbf{x}_{k,i} = \mathbf{x}_{k,j} $\\$ \ \ \ \ \  P(\mathbf{x}_{k,j-1}|\mathbf{z}_{1:k})\leq u_i \leq P(\mathbf{x}_{k,j}|\mathbf{z}_{1:k})$
}\vspace*{0.4cm}
}\vspace*{0.4cm} 
\caption{Bootstrap Particle Filter}
\label{algo:BPF}
\end{algorithm}

\subsubsection{Regularized particle filter}

To deal with the particle degeneracy problem, there exist multiple resampling procedures of diverse complexity. One of these is given by the regularized particle filter, in which the resampling process no longer consists in duplicating particles. The discrete particle population is \textit{regularized} to adopt a continuous form. Each individual particle is assigned a kernel, usually a Gaussian function, with a given bandwidth $h$. The population then takes the form of a \ac{KDE}
\begin{equation}
    p_s(\mathbf{x}) = \sum_{i=1}^N\omega_i \delta(\mathbf{x}-\mathbf{x}_i) \approx \frac{1}{Nh}\sum_{i=1}^{N}K\begin{pmatrix}\displaystyle \frac{\mathbf{x}-\mathbf{x}_i}{h}\end{pmatrix},
\end{equation}
being $K(\cdot)$ a non-negative \textit{window} function, e.g. a Gaussian.
\cite{regularizedpf} discuss on the implementation of regularized particle filters and the design of the kernel estimator. The authors propose to use Gaussian kernels and set the bandwidth according to the sample co-variance of the particle population and the dimension of the state space (cf. \cite{silverman2018density}). This results in an adaptive tuning of the bandwidth, allowing for a more robust resampling step.

\subsection{Markov Chain Monte Carlo}\label{sec:MCMC}

For some applications, there is not a clear definition of the underlying dynamical model, i.e the function $f(\cdot)$ in Eq. \eqref{eq:statediff} cannot be properly characterized. In fact, this is the approach followed for the maneuvering mode of the stochastic hybrid system defined in Eq. \eqref{eq:statediff}, as stated in Section \ref{sec:msotrack}. In these cases, an alternative procedure is given by Markov Chain Monte Carlo (MCMC) methods, which focus on the exploration of a target probability distribution function, disregarding the underlying physical processes. The aim of MCMC algorithms is to sample from the \textit{posterior} distribution rather than actually solving the Bayesian inference problem. According to Bayes' rule (cf. \cite{bayes}), the following assumption holds
\begin{equation}
p(\mathbf{x}_k|\mathbf{z}_{1:k}) \propto p(\mathbf{z}_k|\mathbf{x}_k)p(\mathbf{x}_k|\mathbf{z}_{1:k-1}),
\end{equation}
meaning that the probability of the state $\mathbf{x}$ conditioned on the observation $\mathbf{z}$ is directly proportional to the likelihood of a given state realization $p(\mathbf{z}_k|\mathbf{x}_k)$ multiplied by the probability of the state realization itself $p(\mathbf{x}_k|\mathbf{z}_{1:k-1})$.

Note that in this case the problem is not related to solving Eq. \eqref{eq:FPKEquation} since, in principle, we are not capable of approximating $p(\mathbf{x}_k|\mathbf{z}_{1:k-1})$ following physical process assumptions, i.e. $f(\cdot)$ is not defined.
Instead, one can elaborate an arbitrary \textit{prior}, or \textit{proposal}, distribution $p(\mathbf{x}_k|\mathbf{z}_{1:k-1})$ from which to generate samples that are then retained or discarded according to their likelihood. \ac{MCMC} simulation consists in generating a sequence, or chain, of samples from such proposal distribution. As the length of the chain increases, the sampled distribution converges to the posterior distribution $p(\mathbf{x}_k|\mathbf{z}_{1:k})$.

%There are two general approaches to explore the posterior distribution using \ac{MCMC}: 1) assuming a prior distribution and 2) generating samples based on a single starting point. In the former case, samples are drawn from the prior distribution and its acceptance is based on a likelihood ratio. In the latter, however, a valid approach might be to consider the likelihood a Hamiltonian scalar function, so that each new sample is only dependant on the previous one. In either case, the result is a set (or \textit{chain}) of samples that converge to the desired \textit{posterior} distribution as the number of samples increases. In the following, examples of these two approaches are to be discussed, and its applicability to approximating the Bayesian inference problem in the absence of a valid dynamical model is assessed.

%\subsubsection{Metropolis-Hastings}\label{sec:MH}
\subsubsection{DiffeRential Evolution Adaptive Metropolis (DREAM)}

The \ac{MH} algorithm, first proposed in \cite{metropolishastings}, is one of the simplest and most used Markov Chain Monte Carlo methods. Given a prior \textit{jumping} distribution $\pi(x'|x)$ and an initial point $x_0$, candidate samples $x_{k+1}'$ are subsequently drawn from the \textit{jumping} distribution according to 
\begin{equation}
x_{k+1}'\sim \pi(x_{k+1}'|x_{k}).
\end{equation}
The likelihood ratio 
\begin{equation}
    \alpha = \frac{p(y|x_{k+1}')}{p(y|x_{k})}
\end{equation}
is then compared to a random number $u\sim\mathcal{U}(0,1)$ so that if $\alpha\geq u$ the candidate $x_{k+1} = x_{k+1}'$ is accepted, otherwise $x_{k+1} = x_k$. As the number of samples increases, their distribution converges to the desired \textit{posterior} $p(x|y)$. This is due to the fact that samples are drawn from a jumping distribution, simulating a random walk whose movements are dictated by the ratio of likelihoods. The only tunable parameter within this method is the design of the \textit{jumping} distribution, which in the most naive approximation may be a Gaussian centered on $x_k$ featuring a user-defined covariance matrix.% Note that the magnitude of the covariances dictates the distance travelled by successive candidates, so a proper tuning of these parameters is crucial to avoid frequent travels to low probability regions. The naive \ac{MH} algorithm is not well-suited to high dimensional problems since each candidate comprises an entirely new state realization. Thereby, tuning of the covariance parameters is deemed crucial to draw samples in the support of $p(y|x)$, i.e. samples whose likelihood is not negligible. An alternative can be found in the work by \cite{gibbssampler}, where only a single dimension is varied at each subsequent candidate sample.

%\subsubsection{DiffeRential Evolution Adaptive Metropolis (DREAM)}

There exist multiple alternatives for the definition of the jumping distribution in the %\ac{MCMC} methods based on the
\ac{MH} scheme, tailored to different types of inference problems and target distributions. Among these, one can find the \ac{DREAM} algorithm, developed by \cite{DREAM}. \ac{DREAM} is a multi-chain algorithm, meaning that multiple $x_k$ are updated in parallel to improve convergence and efficiency, especially when exploring multi-modal distributions. Moreover, the parameters of the \textit{jumping} distribution are dynamically adapted to avoid sampling from outside of the posterior support. Successful applications of the \ac{DREAM} algorithm can be found in the works by \cite{DREAM1} and \cite{DREAM2}, where Bayesian inference and parameter uncertainty analyses are applied to complex environmental problems. %Within the current work, \ac{MCMC} sampling is performed using the DREAM algorithm due to its superior capabilities in terms of adaptation to multi-modal posteriors and enhanced convergence.

%In the present work the DiffeRential Evolution Adaptive Metropolis (DREAM) algorithm by J. Vrugt et al. \cite{} is selected, mainly due to its increased performance when compared to naive random-walk $\mathcal{MH}$ algorithms. DREAM uses a Metropolis selection rule to accept or reject samples drawn form a \textit{proposal} distribution. These samples generate a chain or sequence of values that converges to the \textit{target} as this chain is simulated, i.e. more samples are drawn. In fact, DREAM is a multi-chain algorithm, meaning that multiple chains are simulated in parallel to improve convergence and efficiency. The last $n$ values of the different sequences are then incorporated into the population representing the state of the RSO, characterizing the accessible space conditioned on the last observation. Note that multiple hypotheses might stem from this process in case the \textit{posterior} distribution presents a multi-modal shape.

%% file: sections/4_Maneuver_hypothesis_generation.tex
\section{Maneuver hypothesis generation}\label{sec:mhg}

% Within this section we present how we predict the post-maneuver state in the maneuvering mode. In fact, we generate multiple hypotheses, some based on optimal control (from which we may find multiple local minima during sampling) and others based on previously characterized maneuvers (which again may lead to multiple local minima
This section discusses the generation of maneuver hypotheses, which is analogous to characterizing the posterior state distribution $p(\mathbf{x}_k|\mathbf{z}_k,m_k=\mathbb{1})$ when the maneuvering mode is active in the interval $t_{k-1:k}$.
Exploration of this distribution is performed by means of \ac{MCMC} techniques, for which the prior is uniformly distributed and bounded by the admissible control region developed in Section \ref{sec:arc}. Based on the belief of fuel optimal transfers, an approximation for the posterior can be written as
\begin{equation}\label{eq:pp}
    p'(\mathbf{x}_k|\mathbf{z}_k,m_k=\mathbb{1}) \propto p(\mathbf{z}_k|\mathbf{x}_k)\times \exp(-\kappa P(\mathbf{x}_k)),
\end{equation}
where $\kappa$ is a constant parameter and $P(\mathbf{x}_k)$ is the control distance metric defined in Section \ref{sec:control}. The parameter $\kappa$ controls the relative importance of the control cost compared to the measurement likelihood $p(\mathbf{z}_k|\mathbf{x}_k)$, so
it can be set a priori, e.g. based on the control distance corresponding to the centroid of the admissible control region $P(\mathbf{x}^{\star})$ defined in Section \ref{sec:IG}.

Another option is to use previously characterized maneuvers (patterns of life) to approximate the posterior state distribution in the event of a maneuver, i.e.
\begin{equation}\label{eq:ppp}
    p''(\mathbf{x}_k|\mathbf{z}_k,m_k=\mathbb{1}) \propto p(\mathbf{z}_k|\mathbf{x}_k)\times\kappa_h\mathcal{M}(\mathbf{x}_k,t_k).
\end{equation}

$\mathcal{M}(\mathbf{x}_k,t_k)$ in Eq. \ref{eq:ppp} is a \ac{KDE} that contains statistics of the maneuver sequence prior to $t_k$, and $\kappa_h$ is a constant used to control the relative importance of both contributions. These statistics are derived from the relative variation in mean orbital elements implied by the maneuver, as well as the pre-maneuver orbital parameters, as discussed in Section \ref{sec:heuristics}. %Note, in general, heuristics can be defined according to a different criteria as, for example, the operator willingness to perform a maneuver during work hours.

%Sampling from the distributions defined in Eqs. (\ref{eq:pp}-\ref{eq:ppp}) results in the generation of multiple maneuver hypotheses%, whose cardinality $r$ is dictated by the number of modes or local maxima of each posterior
%. These m
Maneuver hypotheses are generated whenever the likelihood of an incoming track $p(\mathbf{z}_k|\mathbf{x}_k)\leq p_{th}$ falls below certain threshold, and they are processed in parallel by means of a regularized particle filter. Whenever the \ac{MLE} corresponding to certain maneuver hypothesis features a sufficiently low measurement likelihood, such hypothesis is pruned and so its assigned particles are eliminated from the sampled distribution representing the state of the target. At some point, the surviving maneuver hypothesis is required to replace the ballistic hypothesis so the filter can detect future maneuvers autonomously. To this end, we keep track of the measurement likelihood of each individual maneuver hypothesis $r$ up to time $t_k$ as
\begin{equation}
    \mathcal{L}(r) = \sum_{j=1}^{k}\phi^{t_k-t_j}p(\mathbf{z}_j|\mathbf{x}_{j}),
\end{equation}
where $\phi\approx 0.95$ is a fading memory factor used to favour more recent tracks. The active ballistic hypothesis at $t_k$ is then indicated by the maximum $\mathcal{L}(r)$, providing a prior orbit to be used in the generation of subsequent maneuver hypotheses.

\input{sections/4_1_control_distance_metric}

\input{sections/4_2_admissible_control_region}

\input{sections/4_3_heuristics_generation}

%% file: sections/4_1_control_distance_metric.tex
\subsection{Control distance metric}\label{sec:control}

%This section is devoted to the development of a simple and efficient control distance metric based on an approximate dynamical model.
We aim to develop an inexpensive metric to characterize the admissible control region as defined in Section \ref{sec:arc}, which can also be used in \ac{MCMC} methods in the form of a log-likelihood. Former definitions of a control distance metric applied to maneuvering space objects can be found in the works by \cite{Holzinger2012} and \cite{Lubey}, which require to determine the solution to an optimal control problem. Therein, the authors emphasize on the scarcity of data for the particular problem of maneuver detection in optical survey scenarios. In the presence of long re-observation times and hidden (unobserved) states, it is convenient to derive a computationally efficient control distance metric that can be used e.g. in statistical sampling methods without significantly increasing the computational cost.

Through a careful simplification of the dynamical model, it is possible to derive metrics that provide a sufficiently close estimation of the control effort % (e.g. proportional to the optimal solution)
required to acquire the post-maneuver orbit from the pre-maneuver one. To this end, we propose to approximate maneuvers as instantaneous velocity changes, or impulsive burns. Let the position and velocity vectors, expressed in Modified Equinoctial Elements, be governed by the set of first order differential equations given in the work by \cite{modifiedEquinoctialElements}, Eq. (9). Hereunder, we will refer to such system of equations as
\begin{equation}
    \frac{d \mathbf{\oe}}{dt} = A(\mathbf{\oe},t)\mathbf{a}_p + \mathbf{b}(\mathbf{\oe},t),
\end{equation}
where the linear dependency of the state derivatives on the perturbing accelerations $\mathbf{a}_p$ has been made explicit for convenience. Under the assumptions that 1) the control input is an impulsive burn, i.e. $\mathbf{a}_p \sim \delta(t-t_M)$, and 2) the sensitivity matrix of the state with respect to the control input, $A$, remains approximately constant between the pre- and post-maneuver orbits, the relative change in orbital elements stemming from an instantaneous velocity variation can be approximated by the linear system
\begin{equation}
\label{eq:linSys}
    \Delta \mathbf{\oe} \approx A(\mathbf{\oe}_0)\Delta V.
\end{equation}
The reader is referred to the former work by \cite{8ECSD} for the complete mathematical derivation.

Transfers between two general orbits require at least two impulsive burns, since single-burn maneuvers lead to a final trajectory that intersects the prior one. At this point, an additional dynamical approximation is introduced: the only accelerating perturbations acting on the dynamical system of interest are those due to the control effort. This assumption of Keplerian motion implies that the spatial geometry of the orbit is invariant in the absence of maneuvers, such that only the location of the object within such orbit (the true longitude $L$) varies in time. The effect of a sequence of $n$ burns in the time-invariant orbital elements $\mathbf{\oe}^i$ can then be approximated as
\begin{equation}
    \Delta \mathbf{\oe}^i \approx \sum_{j=1}^{n_M} A(\mathbf{\oe}^i_{j},L_j)\Delta V_j,
\end{equation}
where the dependency of the sensitivity matrix $A$ on time is implicit through the true longitude at the maneuver epoch $L_j$. Note, however, the intention is to solve the inverse problem, i.e.: given an initial $(\mathbf{\oe}^i_0,L_0)$ and final state $(\mathbf{\oe}^i_f,L_f)$, determine the associated control cost. This boundary value problem is posed in the form of an optimization one, since it is of interest to determine the optimum maneuver sequence in a control cost sense. To simplify the problem, only two impulsive burns are considered, whose joint contribution to the change in orbital elements may be estimated as
\begin{equation}\label{eq:augSys}
    \Delta \mathbf{\oe}^i_e = \begin{pmatrix} A(\mathbf{\oe}^i_0,L_1) & 0 \\ 0 & A(\mathbf{\oe}^i_f,L_2) \end{pmatrix}\begin{pmatrix} \Delta V_1 \\ \Delta V_2 \end{pmatrix}.
\end{equation}
Note the couple of burns $(\Delta V_1, \Delta V_2)$ can be readily solved in a least-squares sense. Let the target change in orbital elements be $\Delta \mathbf{\oe}^i_t$, and the maneuver sequence $\Delta V_{tot} = (\Delta V_1,\Delta V_2)$, then we can define a cost function 
\begin{equation}
    J = \Delta V_{tot}^T\Delta V_{tot} + c_1(\Delta \mathbf{\oe}^i_t - \Delta \mathbf{\oe}^i_e)^T(\Delta \mathbf{\oe}^i_t - \Delta \mathbf{\oe}^i_e),
\end{equation}
being $c_1$ a cost index used to express the relative importance of the control cost with respect to the injection error. Note the subscripts $_t$ and $_e$ are used to indicate the target and estimated changes in orbital elements, respectively. The optimal maneuver sequence $\Delta V_{tot}^*$ that minimizes the cost function can then be obtained by setting the partial derivative of $J$ with respect to $\Delta V_{tot}$ to zero, thereby leading to
\begin{equation}\label{eq:dvmin}
    \Delta V_{tot}^* = \left( 2\left(I + c_1 A'^TA'\right)\right) ^{-1} \cdot 2c_1A'^T\Delta \mathbf{\oe}^i_t
\end{equation}
in which $A'$ is the augmented sensitivity matrix of Eq. \eqref{eq:augSys}, i.e.
\begin{equation}
    A' = \begin{pmatrix} A(\mathbf{\oe}^i_0,L_1) & 0 \\ 0 & A(\mathbf{\oe}^i_f,L_2) \end{pmatrix}.
\end{equation}

Despite the solution given by Eq. \eqref{eq:dvmin} simply requires a matrix inversion, the following constrained non-linear optimization problem needs to be solved to determine the optimum pair of true longitudes $(L_1,L_2)$

\begin{equation}\label{eq:ocpL}
\begin{array}{ll}
\textrm{Minimize} & J =  \Delta V_{tot}^T\Delta V_{tot} + c_1(\Delta \mathbf{\oe}^i_t - \Delta \mathbf{\oe}^i_e)^T(\Delta \mathbf{\oe}^i_t - \Delta \mathbf{\oe}^i_e) \\
\textrm{subject to:} & L_0\leq L_1 \leq L_2 \leq L_f
\end{array}
\end{equation}

Compared to the control distance metric proposed by \cite{Holzinger2012}, the size of the \ac{NLP} problem is dramatically reduced in this case. A practical approach to solve the optimization in Eq. \eqref{eq:ocpL} may consider the use of a gradient descent method, which can be applied to an appropriate set of initial conditions to improve convergence to a global minimum. 

Finally, the definition of the control distance metric $P$ reads
\begin{equation}
    P = ||\Delta V_1^*||+|| \Delta V_2^* ||,
\end{equation}
where $||\cdot||$ is used to indicate the Euclidean norm of a vector, and $(\Delta V_1^*, \Delta V_2^*)$ are the first and second burns that stem from evaluating Eq. \eqref{eq:dvmin} at the optimum set of true longitudes $(L^*_1,L^*_2)$. The metric $P$ has therefore units of velocity, as it is the result of solving the fuel optimal two-burn transfer between two Keplerian orbits under certain linear approximations.

%% file: sections/4_2_admissible_control_region.tex
\subsection{Admissible control region}\label{sec:arc}

% Link this section with MCMC and say that the prior or jumping distribution is built based on the admissible control region...
In order to apply \ac{MCMC} methods to estimate the state transition in maneuvering intervals, it is convenient to define proper bounds for a prior distribution of the state. In the following, we take advantage of the work by \cite{IODShortArc}. who propose a revisit of the \textit{admissible region} developed by \cite{milaniAR} tailored to the needs of \ac{RSO} cataloging. Therein, bounds for the expected range and range-rate values $(\rho,\dot{\rho})$ are developed based on different orbital regimes. This is, given an observation in the attributable format $(\alpha,\delta,\dot{\alpha},\dot{\delta})$, the position and velocity of a target object is completely defined by the position and velocity of the observing site $(\mathbf{r_s},\mathbf{\dot{r}_s})$ and the range and range-rate $(\rho,\dot{\rho})$ according to the expressions in Eqs. (\ref{eq:rho}-\ref{eq:drho}).

It is then possible to define bounds on the range and range-rate given admissible sets for the semi-major axis $a$ and eccentricity $e$. This method is applicable to the association of two observations, where there is not a clear definition of the orbital state of the observed object. In the case of maneuver detection, the association problem is defined between a clearly established orbit (of a cataloged object) and a hypothetical post-maneuver observation. Direct application of the \textit{constrained admissible region} as defined by \cite{IODShortArc} can be used to reduce the search space for post-maneuver state estimation, but does not consider the underlying physical process undergone by the object.

Hereunder, we embrace an alternative definition, termed the \textit{admissible control region}, in which the space accessible to the object is not bound in terms of semi-major axis and eccentricity, but based on the control effort required to reach certain range and range-rate values. A similar approach has already been explored by \cite{SERRA2021}, who propose a convexification of the \textit{admissible region} based on a control-related energy metric. Still, bounds for the accessible space are given in terms of the maximum variation in this energy metric, largely based on the former control distance metric by \cite{Lubey}. These bounds are then translated into maximum and minimum semi-major axis and eccentricity values, thus disregarding the relative geometry of the initial and final orbits in terms of angular distance. This approach is beneficial from the association standpoint as there is no need to solve a \ac{NLP} problem to accept or reject the maneuver hypothesis in a preliminary step.

The interest of the present work is not to tackle the maneuver detection and data association considering multiple-targets as in \cite{SERRA2021}, but to emphasize on the single-target case. Thereafter, we propose a target-based \textit{admissible control region}
\begin{equation}\label{eq:acr}
    \mathcal{C}(\mathbf{x}) = \{ \mathbf{x} :  h(\mathbf{x})=\mathbf{z}, P(\mathbf{x})\leq P_{adm} \}
\end{equation}
in which an admissible set for the post-maneuver range and range-rate is determined in terms of a maximum expected control effort $P_{adm}$. Attending to the topology of the control distance metric derived in Section \ref{sec:control}, upper and lower bounds for the range and range-rate are elaborated.

\subsubsection{Centroid of the admissible region}\label{sec:IG}

% Here we will argue that we need a starting point, a sort of centroid for the admissible control region
% In order to determine this point we need to perform some sort of optimization in range and range rate over the L1-L2 metric space, which might be costly...
% And not only that, the topology might be non-smooth so gradient based search may only be used up to a certain point, given the initial guess is sufficiently close to a local minimum.
% In fact we do not want to find a global optimum since it cannot be demonstrated that the global optimal maneuver than is able to completely fit an observation gives us the best post-maneuver state estimate -> IT CANNOT BE DEMONSTRATED BTW
% Then we just want to find some reasonable starting point to determine our reasonable set (or admissible region)

In contrast to the former definition of the admissible region, which yields a symmetric set in the range-rate space, our proposal requires a starting point. This point can be thought of as the \textit{centroid} of the admissible set. An approximation to the latter may be given by the state compliant with an \textit{attributable} that is closest to the pre-maneuver orbit $\mathbf{x}_b$ in terms of the metric $P$, and is defined as

\begin{equation}\label{eq:x*}
\mathbf{x}^{opt}=\{ \mathbf{x} : h(\mathbf{x})=\mathbf{z}, \underset{\mathbf{x}}{\text{argmin}}P(\mathbf{x}) \}.
\end{equation}

\begin{figure}[!htb]
    \centering
    \includegraphics[scale = 0.5]{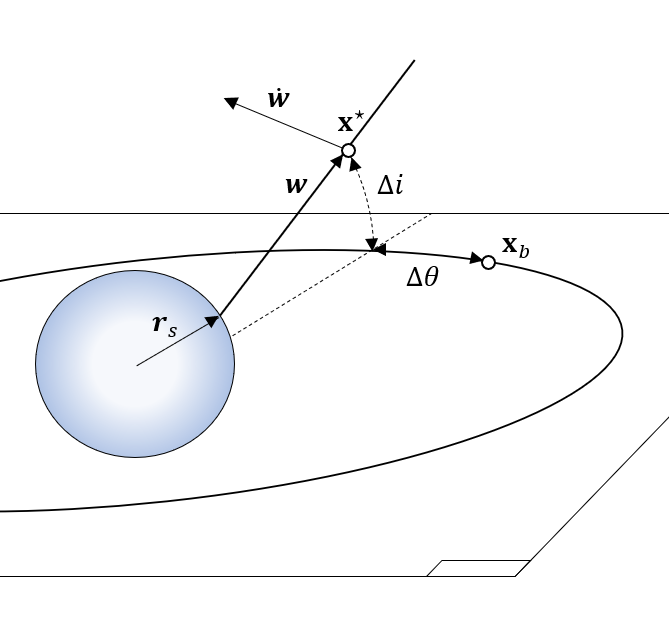}
    \caption{Geometrical definition of the \textit{centroid} $\mathbf{x}^{\star}$ in terms of the pre-maneuver orbit $\mathbf{x}_b$ and the observables $\mathbf{z}\equiv(\mathbf{w},\dot{\mathbf{w}})$.}
    \label{fig:igsketch}
\end{figure}

The determination of this point involves solving an optimization problem over the range and range-rate in order to find the global minimum of the control distance metric. Note, in general, such point may not provide an accurate estimate of the true post-maneuver state since the actual purpose of a maneuver is to acquire a final orbit; not an optical track. Thereafter, and in the interest of computational efficiency, we propose to use an approximation to this global minimum $\mathbf{x}^{\star}\approx\mathbf{x}^{opt}$ solely based on geometrical considerations. It is, in essence, a valid \textit{initial guess} to be used in the former optimization problem. Figure~\ref{fig:igsketch} illustrates the definition of $\mathbf{x}^{\star}$, which can also be expressed as
\begin{equation}\label{eq:xstar}
    \mathbf{x}^{\star} = \left\{ \mathbf{x}(i,\theta):  \frac{\partial}{\partial(i,\theta)}\left((\mathbf{z}-h(\mathbf{x}))^T(\mathbf{z}-h(\mathbf{x}))\right)=0 \right\},
\end{equation}
being $i$ and $\theta$ the inclination and true anomaly of the orbit, respectively. This \textit{centroid} is a low-dimensional transformation of the pre-maneuver orbit, involving the minimum number of orbital parameters required to match the observation.

\begin{figure}[!htb]
    \centering
    \includegraphics[trim = 10 10 20 10, clip,width = .7\linewidth]{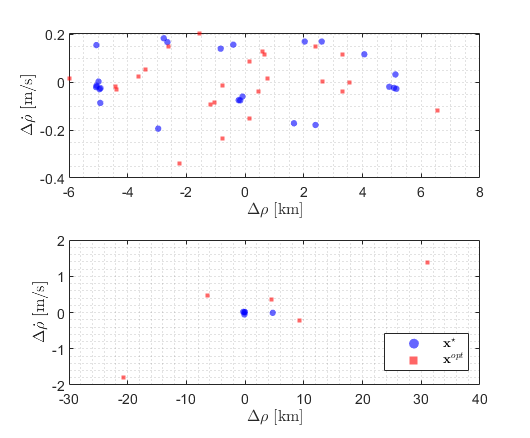}
    \caption{Range and range rate differences with respect to the true post-maneuver state $(0,0)$ for a set of east-west (top) and north-south (bottom) maneuvers.}
    \label{fig:centroid_val}
\end{figure}

% Add some figure comparing xocp, xopt and xstar
Fig. \ref{fig:centroid_val} compares the post-maneuver estimation accuracy for the optimal state in terms of the control distance metric $\mathbf{x}^{opt}$ and the proposed centroid $\mathbf{x}^{\star}$ for two different maneuver types. Albeit small, the fuel optimal estimate shows an increase in accuracy when estimating east-west maneuvers since they usually imply changes in semi-major axis and/or eccentricity. These are not accounted for in the definition of the centroid, and so errors in range and range rate are mostly due to variations in those orbital parameters. On the contrary, the estimate given by the centroid for north-south maneuvers is significantly closer to the true post-maneuver state than that of $\mathbf{x}^{opt}$. The latter supports the fact that a fuel optimal transfer from an initial orbit to a single optical observation does not necessarily provide a good estimate. In this case, since out-of-plane maneuvers require higher impulses than in-plane ones, $\mathbf{x}^{opt}$ overestimates the relative change in semi-major axis and eccentricity, with the aim of minimizing the required inclination change to match the observation. Based on this analysis, approximating the centroid of the admissible region as $\mathbf{x}^{\star}$ seems a good trade off between computational efficiency and accuracy, and it is indeed compliant with an orbit preserving assumption, in the sense that one only expects phasing ($\Delta \theta$) and inclination ($\Delta i$) change maneuvers. 

\subsubsection{Admissible region topology}\label{sec:artopo}

An analysis of the behavior of the control distance metric in the $\rho$-$\dot{\rho}$ space is required in order to define proper bounds for the admissible control region. In general, one can define these bounds in the form of an upper energy level, as suggested in Eq. \eqref{eq:acr}. Therein, the maximum admissible control effort $P_{adm}$ may be defined in absolute and relative terms as
\begin{equation}\label{eq:padm}
    P_{adm} = min(P_{max},max(P_{min},k_PP(\mathbf{x}^{\star}))).
\end{equation}

This allows to reduce the search space when observations are close to the pre-maneuver orbit, through $k_P$ and $P_{min}$, but also when $P(\mathbf{x}^{\star})$ approaches $P_{max}$. More complex thresholding functions can be elaborated, yet the simplistic approach given by Eq. \eqref{eq:padm} provides reasonable bounds at a modest tuning effort.

\begin{figure*}[!htb]
    \centering
    \begin{multicols}{2}
    \hspace*{-0.7cm}
    \includegraphics[width = 1.15\linewidth]{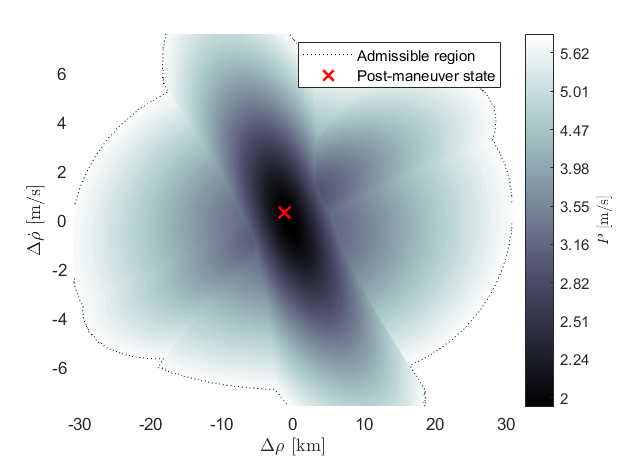}
    \hspace*{-0.7cm}
    \includegraphics[width = 1.15\linewidth]{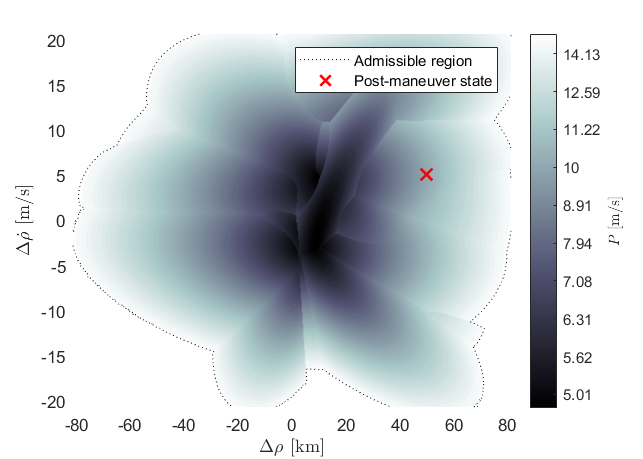}
    \end{multicols}
    \caption{Admissible control region for a combined north-south/east-west maneuver (left) and an orbit raising maneuver (right), expressed in relative $(\rho,\dot{\rho})$ values with respect to $\mathbf{x}^{\star}$. The true post-maneuver state is indicated with a red cross, while the outer boundaries of the admissible set (where $P_{max} = 15$ m/s, $P_{min} = 3$ m/s and $k_P = 3$) correspond to the black dotted line.}
    \label{fig:admissibleregion}
\end{figure*}

Figure~\ref{fig:admissibleregion} depicts the behavior of the control distance metric $P$ in the range and range-rate space for two different maneuvers, centered at $\mathbf{x}^{\star}$. It can be seen that $\mathbf{x}^{\star}$ provides a good approximation to the \textit{centroid} of the admissible region in both cases, and it is found to be close to the optimum in terms of control distance defined in Eq. \eqref{eq:x*}. The topology of the control metric $P$ exhibits a non-smooth behavior between local valleys, potentially hindering the application of gradient-based methods for determining $\mathbf{x}^*$. This behavior is thought to be related with the assumptions made in the definition of the control metric $P$, especially the one in Eq. \eqref{eq:linSys}.

The \textit{centroid} $\mathbf{x^{\star}}$ is uniquely defined in the unobservable range and range-rate space by $(\rho^{\star},\dot{\rho}^{\star})$, since the observed magnitudes are expected to match the measurements. In order to approximate the complicated outer boundaries of the admissible control region, depicted in Fig. \ref{fig:admissibleregion}, one can define certain search directions in the unobservable space along which to determine the intersections with $P_{adm}$. For the two cases shown, a polytopic approximation may provide accurate results given a proper selection of search directions. Accordingly, the authors propose to approximate the admissible control region as an orthotope with search lines parallel to the range and range-rate axes and passing through the \textit{centroid} $\mathbf{x^{\star}}$, in the form of an axis-aligned minimum bounding box. Moreover, the orthotope that numerically represents the admissible control region is extended in the observable space by 3-$\sigma$ bounds along each dimension to accommodate measurement uncertainty.

\begin{figure*}[!htb]
    \centering
    \hspace*{-0.6cm}
    \includegraphics[trim = 60 60 60 10, clip,width = 1.1\linewidth]{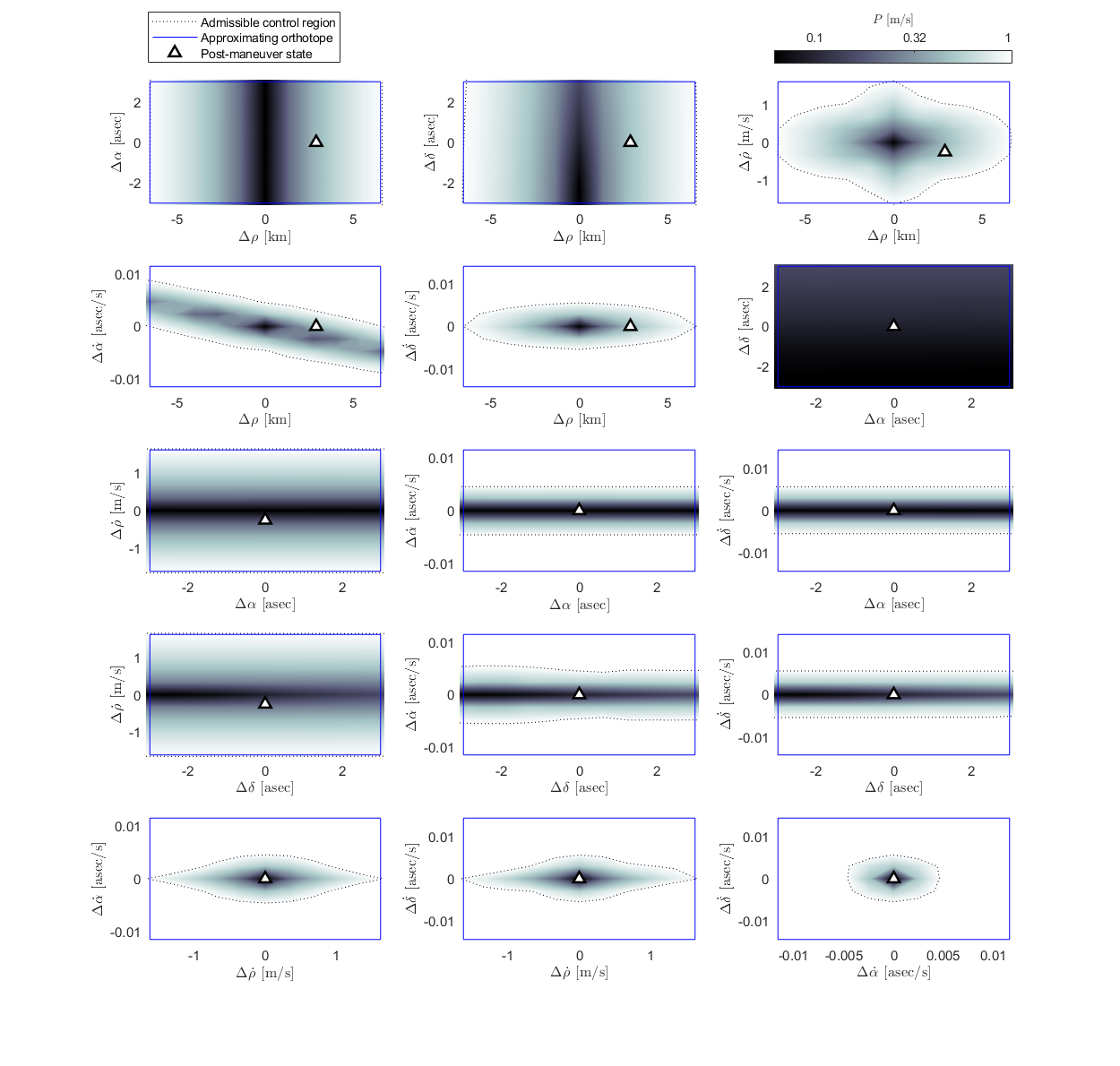}
    \caption{Topology of the admissible control region for an east-west maneuver in the observable $\left( \alpha , \delta , \dot{\alpha} , \dot{\delta} \right)$ and unobservable $\left( \rho, \dot{\rho}\right)$ spaces. The control distance metric for the centroid is $P(\mathbf{x}^{\star})=0.06$ [m/s], while that of the post-maneuver state is $P(\mathbf{x}^+)=0.22$ [m/s]. The distance between the centroid and post-maneuver states in the unobservable space is $\Delta \rho = 2.90$ [km] and $\Delta\dot{\rho} = -0.254$ [m/s].}
    \label{fig:acr_orthotope}
\end{figure*}

An example of such polytopic approximation can be consulted in Fig. \ref{fig:acr_orthotope}, where the admissible control region is depicted in topocentric spherical coordinates for an east-west maneuver. Therein, two-dimensional maps of the control metric are shown for all the possible combinations of states, centered at $\mathbf{x}^{\star}$. The distribution of $P$ within the $\alpha$-$\delta$ admissible space is nearly isotropic, showing control metric values close to the minimum, thus indicating that the boundaries implied by the measurement uncertainty are much lower that those given by the maximum expected control effort. On the contrary, the behavior of $P$ in the $\dot{\alpha}$-$\dot{\delta}$ plane feature higher gradients that eventually result in a over-estimation of the admissible control region if the expected measurement noise is used (determined according to \cite{covariancered}, Eq. 8). The majority of projections show a single global minimum with iso-$P$ lines that are either concentric or parallel to one of the axes. That is not the case for the $\rho$-$\dot{\alpha}$ plane, showing multiple local minima and some correlation between the range and the right ascension rate.
The assumption $\mathbf{x}^{opt}\approx \mathbf{x}^{\star}$ seems to hold or at least $\mathbf{x}^{\star}$ appears to be closer to $\mathbf{x}^{opt}$ than the true post-maneuver state. Approximating $\mathcal{C}(\mathbf{x})$ as an orthotope following an axis-aligned box may result in an overestimation of the space accessible in terms of $P$, as indicated by the white regions outside the dotted line boundaries of Fig. \ref{fig:acr_orthotope}. Nonetheless, it provides an extremely efficient computation of the expected region wherein to search for the post-maneuver state, which is desirable since $\mathcal{C}(\mathbf{x})$ must be determined individually for any post-maneuver observation. In fact, a more precise definition of the boundaries would not yield any increase in estimation accuracy since every candidate point within the admissible region is weighted according to its associated control distance.

% Justify the incorporation of figure 5, which is huge. Comment on how the control metric limits the accessible space for the unobserved variables but also how it can be used to narrow the search for observed variables also, i.e. range rates. Maybe comment on the complexity implied by the dimensionality of the system and how the different views/sections/cuts incorporate knowledge regarding the maneuverability of the object. Also we should comment that, in principle, the noise in right ascension and declination is so low that we could reach any value at the cost of x_opt....

%% file: sections/4_3_heuristics_generation.tex
\subsection{Heuristic characterization}\label{sec:heuristics}

% Within this section we will briefly introduce how we perform maneuver characterization based on the comparison of initial and final orbits. Then we will briefly talk about gaussian mixture clustering and how we can use the maneuver history in the form of a gaussian mixture to perform conditioned sampling on the admissible region

It is common practice to approach the maneuver detection and estimation problem using the maneuver history or patterns of life, see for instance \cite{siminski2017correlation,shabarekhML,abay2018maneuver}. These works propose the use of heuristic methods, based on a statistical characterization of the maneuver history and \ac{ML} techniques. In general, spacecraft maneuver to be kept inside a specific orbital slot, usually referred to as station-keeping. Dynamical perturbations acting on the spacecraft motion are typically well-characterized, so that the various station-keeping maneuvering modes are already devised at the mission design stage. These usually feature certain temporal frequency and similar control magnitudes, with the aim of simplifying the operational workload. Thereafter, application of heuristics and \ac{ML} methods seems suitable to this scenario, and has seen successful implementations in the previously cited works.\\

The proposal herein suggested is largely based on the work by \cite{siminski2017correlation}, in which maneuvers are characterized in terms of relative variations in certain orbital elements $\xi = [\Delta a \ \Delta e \ \Delta i]^T$.
As discussed in Section \ref{sec:fshs} state space filtering is performed sequentially by means of a regularized particle filter. Post-maneuver observations trigger the generation of maneuver hypotheses, which are tested for correlation with subsequent tracks. At some point, there is a single hypothesis or group of equivalent hypotheses that survive, and then the filter is assumed to have converged in terms of state (and mode) estimation accuracy, i.e. there is a change in the active ballistic hypothesis indicated by $\mathcal{L}(r)$. It is not until such convergence is detected that the regularized particle filter is run in reverse mode (\cite{BSMC}), and the last (earliest) observation that associates with the surviving hypothesis is deemed the first post-maneuver track. Immediate pre- and post-maneuver orbits are then compared in terms of the relative variation in orbital elements. Note that statistics are readily obtained from this comparison since the pre- and post-maneuver states are given in the form of sampled distributions. %For every pair of pre- and post-maneuver orbits, a \ac{KDE} $\mathcal{M}(\xi,t_k)$ is augmented with $\hat{\xi}_k=\mathbb{E}[\xi_k]$ and $\Xi_k=\text{Cov}[\xi_k]$ as
Detected maneuvers are thus represented by a \ac{KDE} $\mathcal{M}(\mathbf{x}_k,t_k)$ that is continuously updated with every $\hat{\xi}_k=\mathbb{E}[\xi_k]$ and $\mathbf{\Xi}_k=\text{Cov}[\xi_k]$ as
\begin{equation}
    \mathcal{M}(\mathbf{x}_k,t_k)=\frac{1}{n_M}\sum_{j=1}^{n_M}\frac{\exp{\left[-\frac{1}{2}(\xi-\hat{\xi}_j)^T\mathbf{\Xi}_j(\xi-\hat{\xi}_j)\right]}}{\sqrt{(2\pi)^{n_{\xi}}|\mathbf{\Xi}_j|}},
\end{equation}
being $n_M$ the number of detected maneuvers up to time $t_k$, and $n_{\xi}$ the dimensionality of the \textit{feature} vector $\xi$. In this case we have adopted a multivariate normal kernel estimator due to its simplicity, but there are other alternatives based on e.g. automatic bandwidth selection (see \cite{wand1994multivariate}).

%As maneuvers are detected, they are characterized and incorporated into a \ac{KDE} $\mathcal{M}(t_k)$ that can then be used to approximate future maneuvers, e.g. shaping the admissible control region defined in Section \ref{sec:arc}. The latter can improve the convergence of statistical sampling methods to approximate the post-maneuver state distribution, as the number of samples corresponding to high probability density regions is expected to increase, i.e. the narrower the sampling space the more efficient the sampling procedure.\\

The techniques discussed in Section \ref{sec:MCMC} are used to explore the post-maneuver state distribution conditioned on both the incoming observations and the maneuver history, i.e. sample from $p\left( \mathbf{x}_k|\mathbf{z}_k, \mathcal{M}(\mathbf{x}_k,t_k) \right)$. In doing this, samples generated from the admissible control region $\mathcal{C}(\mathbf{x})$ are evaluated in terms of $[\Delta a \ \Delta e \ \Delta i]$, and this relative change is compared against the \ac{KDE} containing the information related to previously characterized maneuvers. Thereafter, changes in orbital elements that are compliant with previous maneuvers are favoured, potentially leading to a more precise post-maneuver state recovery in the presence of a repetitive maneuver plan.

%% file: sections/5_results_and_comparison.tex
\section{Results and comparison}\label{sec:results}

%Analysis of the proposed approach is provided in the following, both from the tracking and maneuver detection perspectives. Comparisons are drawn against former methodologies, in particular, a moving horizon estimator implementation% and a Multi-Hypothesis Unscented Kalman filter
%. Results are obtained for synthetic data, representative of a real operational scenario. Performance is evaluated in terms of maneuver detection metrics and post-maneuver state estimation (tracking) accuracy, which leverages diversion from ground truth and a consistent uncertainty characterization.

\input{sections/5_1_metrics}
\input{sections/5_2_methods}
\input{sections/5_3_synthetic}

%% file: sections/5_1_metrics.tex
\subsection {Performance metrics}\label{sec:metrics}

% should we give details for every single metric we use?
% first we should talk about the Cramer Rao Lower bound as a measure of the volatility and the Normalized Estimation Error Squared as a measure of the biasedness...
% In fact, it makes more sense to me to present those plots with the NEES distribution for the entire tracking window, since they summarize the consistency of the filter (jointly mixing accuracy and proper uncertainty characterization).

%Hereunder, we define some metrics used to infer the tracking performance of the proposed method.
Two different metrics have been devised to characterize the response of the algorithm in terms of accuracy and consistency. While the former is directly affected by the signal to noise ratio, a statistically consistent (unbiased) filter should be able to adjust its considered uncertainty in order to deliver unbiased estimates.

The objective of the proposed filtering scheme is to jointly solve the maneuver detection and tracking problem. Thereafter, emphasis should be in the post-maneuver state estimation and so the root-mean-square error with respect to the true post-maneuver state $\mathbf{x}^+$ is given as a function of the number of tracks elapsed after each maneuver, $n_T$, as defined in Eq. \eqref{eq:rmse}. Therein, $m$ is the active mode introduced in Eq. \eqref{eq:mode} and $\mathbb{1}_1=[1 \ 0 \  ... \ 0]$ is the indicator function with dimension $n_T$.

\begin{equation}
    \textit{RMSE}_{\mathbf{x}}(n_T) = \sqrt{ \frac{\sum_{i=1}^{N_i} (\mathbf{x}_i - \mathbf{x}^+_i)^2 }{N_i} }, \ \ \ \ \forall i : m_{i-n_T:i} = \mathbb{1}_1
    \label{eq:rmse}
\end{equation}

Besides, with the aim of providing a means to analyze the uncertainty realism, or statistical consistency of the filter, the \ac{PCRB} is used as proposed in \cite{PCRB}. The \ac{PCRB} is an analogy of the Cramér Rao Lower Bound (CRLB) (see \cite{Rao1992}) when applied to the estimation process of random parameters. The \ac{PCRB} can be regarded as a lower bound for the variance of any unbiased estimator, and is given by the inverse of Fisher's information matrix
\begin{equation}
    \mathcal{J}_{ij} = -\mathbb{E} \left\{ \frac{\partial^2 logp(\mathbf{x},\mathbf{z})}{\partial x_i \partial x_j } \right\}.
\end{equation}
A recursive derivation of the Fisher information matrix, $\mathcal{J}$, for the particular case of Gaussian process and measurement noise $g(\cdot)\mu_k$, $q(\cdot)\gamma_k$ with co-variance $Q_k$ and $R_k$, is given by Eq. \eqref{eq:fim} for the non-linear discrete stochastic process described in Eqs. \eqref{eq:statediff} and \eqref{eq:zofx}. The expectation operators in Eqs.~(\ref{eq:d11}-\ref{eq:d22}) are approximated by means of Monte Carlo averages since the dynamical and measurement models are, in general, non-linear.

\begin{align}
\label{eq:fim}
    \mathcal{J}_{k+1} & = D^{22}_k - D^{21}_k(\mathcal{J}_k+D^{11}_k)^{-1}D^{12}_k \\
    \mathcal{J}_{0,ij} & = -\mathbb{E} \left\{ \frac{\partial^2 logp(\mathbf{x}_0)}{\partial x_i \partial x_j } \right\} \\ \label{eq:d11}
    D^{11}_k & = \mathbb{E}\left\{ \left[ \nabla_{\mathbf{x}_k} f^T(\mathbf{x}_k) \right]Q_k^{-1}\left[ \nabla_{\mathbf{x}_k} f^T(\mathbf{x}_k) \right]^T  \right\}\\
    D^{12}_k & = -\mathbb{E}\left\{ \nabla_{\mathbf{x}_k} f(\mathbf{x}_k)^T\right\}Q_k^{-1}     \ \ \ \ \ \ \ \ \ \ \ D^{21}_k = \left[D^{12}_K\right]^T\\
    D^{22}_k & = Q_k^{-1}+\mathbb{E}\left\{ \left[ \nabla_{\mathbf{x}_{k+1}} h^T(\mathbf{x}_{k+1}) \right]R_k^{-1}\left[ \nabla_{\mathbf{x}_{k+1}} h^T(\mathbf{x}_{k+1}) \right]^T  \right\}\label{eq:d22}
\end{align}

Note this derivation is only valid for the particular case of known active mode $m_k=\emptyset$. In order to simplify the approach followed to compute the PCRB, 1) the active mode between subsequent observations will be considered known, and 2) the dynamical process noise $Q_k$ when $m_k=\mathbb{1}$ will be assumed equal to the co-variance of the distribution $p\left(\mathbf{x}_k|\mathbf{z}_k,\mathcal{M}(\mathbf{x}_k,\mathcal{T})\right)$, with $\mathcal{M}(\mathbf{x}_k,\mathcal{T})$ a KDE representing the entire set of true maneuvers characterized as described in Section \ref{sec:heuristics}.

Orbital uncertainty evolution in Cartesian co-ordinates presents a highly non-linear behavior, unlike parametric representations such as classical orbital elements or \ac{MEE}, among others (cf. \cite{understandingUnc}). Thereafter, uncertainty is to be characterized in \ac{MEE} and compared to the aforementioned \ac{PCRB} to test the statistical efficiency of the proposed filter. 

In this regard, we can construct a distance metric analogous to the Mahalanobis distance (see \cite{mahalanobis})
\begin{equation}\label{eq:d2_dist}
    d_k^2 = (\mathbf{x}_k-\hat{\mathbf{x}}_k)^TC(\mathbf{x}_k-\hat{\mathbf{x}}_k)
\end{equation}
where $\hat{\mathbf{x}}_k$ is the reference or ground truth value for the state estimate and $C=\left\{\mathcal{J}_k, \ \Sigma_k^{-1}\right\}$ may be the PCRB or the co-variance matrix of the estimates. The above distance $d^2\sim\chi^2(n_x)$ follows a \textit{chi-square} distribution provided the random vector $\mathbf{x}$ is normally distributed. An unbiased and statistically efficient method would comply with the latter for $C=\mathcal{J}_k$, whereas a consistent uncertainty characterization would feature $d^2\sim\chi^2(n_x)$ for $C=\Sigma_k^{-1}$. If the distribution of the estimation error distance $d^2$ features a lower skewness than $\chi^2(n_x)$ then the uncertainty is overestimated and the filter can be considered pessimistic. On the contrary, a higher skewness is an indicative of an underestimated uncertainty (optimistic filter) and possibly a biased estimation.

%% file: sections/5_2_methods.tex
\subsection{Benchmarking methods}\label{sec:methods}

Various filter implementations have been evaluated in order to analyze the performance and improvements of the proposal over a standard operational approach. The differences between these methods are mainly related to the filtering scheme, maneuver detection and post-maneuver state estimation, so environmental modelling and measurement association remain the same.

With regard to the former, the dynamical model used for state estimation have the following characteristics:
    \begin{itemize}
        \item Non-spherical Earth of degree and order 10.
        \item Third-body perturbations of Sun and Moon.
        \item Cannonball model for the \ac{SRP} with a conical solar and lunar eclipse model, using fraction of illumination for penumbra regions.
    \end{itemize}
    
Measurement association is dictated by the Mahalanobis distance from the state to the observation,
\begin{equation}
    d'^2_k = (\mathbf{z}_k-h(\mathbf{x}_k))R_k^{-1}(\mathbf{z}_k-h(\mathbf{x}_k)),
\end{equation}
which is assumed to follow a $\chi^2$ distribution with 4 degrees of freedom. The no-maneuver association threshold, $p_{th}$ is defined as the 3-$\sigma$ gate of the aforementioned distribution for the current study. Post-maneuver association is subject to a threshold set on the maximum expected control effort, so that if the expected value of the post-maneuver state is farther in terms of $P$, i.e. $P \geq P_{max}$,  then the track remains uncorrelated. This maximum expected control effort is highly related with the definition of the admissible control region $\mathcal{C}(\mathbf{x})$, whose outer boundaries are set attending to the following thresholds: $P_{max}=10$ m/s, $P_{min} = 1$ m/s and $k_{P}=3$.

The baseline method, assumed an operational standard, consists in a moving horizon estimator (hereafter \textbf{MHE}) that considers up to 6 subsequent tracks. Measurement association and maneuver detection is based on the aforementioned thresholds, also considering the admissible region defined using the developed control distance metric. %Estimation is re-initialized after a maneuver is detected (no-association)
Post-maneuver state estimation is conditioned on the sequence of observations after the detected maneuver (no-association), and the loss function is augmented to include the control distance $P$ to the previous orbit, so that the post maneuver estimate corresponds to $\mathbf{x}^{opt}$. As a benchmarking option, the same moving horizon estimator is used, but this time considering the true maneuver sequence in spite of detecting a maneuver in terms of the Mahalanobis distance between the track and the ballistic trajectory. The latter method is termed \textbf{MHE II}, and is assumed to deliver the best estimate in a Bayesian sense when no \textit{heuristic} or \textit{a priori} maneuver information is included.

The proposed method consists in a regularized particle filter with a variable population size, wherein multiple maneuver hypotheses may be active. Each of these maneuver hypotheses is associated a fixed number of particles, $N_H = 1000$, which are individually tested for association. A former implementation, termed \textbf{SHF}, discards the use (and automatic generation) of heuristics, therefore solely relying on the control distance metric to approximate maneuvers. The complete proposal is realized through \textbf{SHF II}, where heuristics are automatically generated based on detected maneuvers (in a feedback loop), in principle using all the available information in the estimation process through an alternative sampling conditioned on previous maneuvers.

%% file: sections/5_3_synthetic.tex
\subsection{Synthetic measurements}\label{sec:res_synthetic}

A simulation is carried out for a \ac{GEO} spacecraft performing station-keeping maneuvers, and the test scenario is defined as follows:
\begin{itemize}
    \item \emph{Subject:} \ac{GEO} Spacecraft equipped with chemical propulsion. Its assigned orbital slot comprises a mean longitude band $\ell = -4.8 \pm 0.2 \degree$ and an inclination band $i = 2\pm 0.05 \degree$. The object is simulated for a total duration of 401 days (03/09 - 05/10) using a dynamical model including the following perturbations:
    \begin{itemize}
        \item Non-spherical Earth of degree and order 70.
        \item Third-body perturbations of Sun, Moon and Planets (including Pluto).
        \item Cannonball model for the \ac{SRP} with a conical solar and lunar eclipse model, using fraction of illumination for penumbra regions.
        \item Dynamical noise is introduced in the SRP coefficient $B$ in the form of random Poisson temporal variations with parameter $\lambda = 7$ days and magnitude $\Delta B \sim \mathcal{N}(0,10^{-2})$.
        \item Solid Earth and ocean tides.
        \item General Relativity.
    \end{itemize}
    \item \emph{Optical Sensor Network:} two optical ground telescopes located at Zimmerwald (AIUB Zimmerwald's Observatory) and Tenerife (ESA Optical Ground Station). The optical survey presents the following characteristics:
    \begin{itemize}
    \item Elevation mask of $20 \degree$.
    \item Solar phase angle between $0 \degree$ and $90 \degree$.
    \item Angular distance to Earth shadow $\theta > 0 \degree$.
    \item Observation model for both right ascension $\alpha$ and declination $\delta$ featuring a zero-mean Gaussian noise with standard deviation $\sigma_{\alpha,\delta}=1^{\prime\prime}$.
    \item Mean re-observation time of two days for each individual sensor.
    \item Track length $T\sim \mathcal{U}(2,10)$ min.
    \item Tracks are reduced to the \textit{Attributable} format by performing a linear regression with respect to the mean epoch.
    \item Observation covariance is determined according to the time span and number of observations of a given track as suggested in \cite{covariancered}, Eq. (8), for a second order fit on the sequence of $\alpha$-$\delta$ pairs.
    \end{itemize}
\end{itemize}

A total number of 20 detectable maneuvers, consisting of 29 detectable burns, are distributed along 401 days and 362 tracks. 5 of them are single-burn NSSK, while the total number of (double-burn) EWSK is 16. Note that for these maneuvers (and burns) to be detectable an observation needs to be obtained between them, otherwise two or even three burns may be collapsed into one distinguishable orbit update. Table \ref{tab:mandet} gathers the results for two alternative methods: the moving horizon estimator (MHE) and the proposed combined optimal-heuristic hybrid filter (SHF II). The latter 
has shown more accurate in detecting maneuvers (and burns) at the correct epoch, being able to properly identify all detectable maneuvers and an additional burn. The MHE filter is also capable of identifying all the detectable maneuvers but four of them are detected with a delay of one observation. Nonetheless, the higher sensitivity of SHF II triggered two false detections. Both are found to be caused by an overestimation of the eccentricity, which leads to a bad characterization of the mean longitude for an EWSK maneuver and a lower estimated inclination in the case of a NSSK maneuver. On the contrary, since the MHE is capable of performing significant orbit updates, no false detections have been reported.

\begin{figure}[!htb]
    \centering
    \includegraphics[trim = 15 0 20 120, clip, width = .73\textwidth]{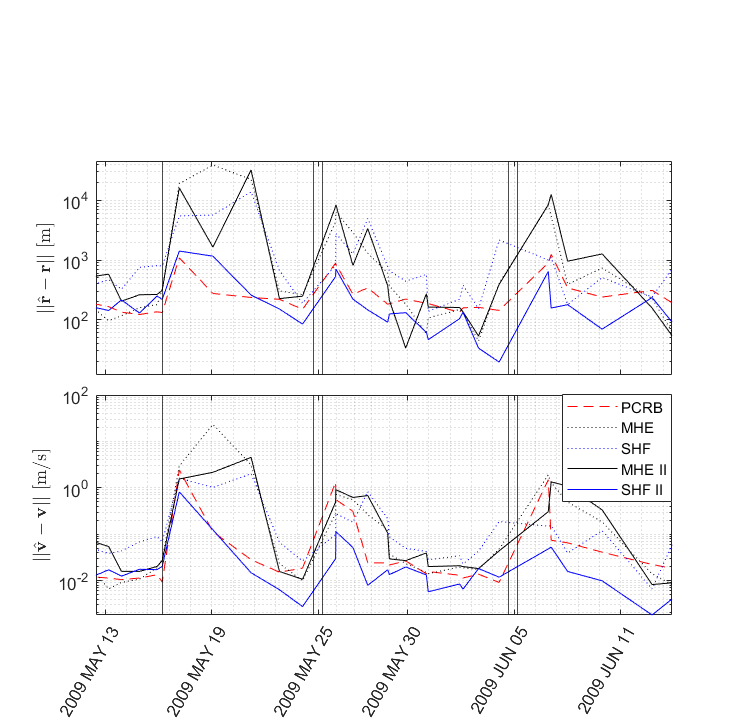}
    \caption{Position (top) and velocity (bottom) estimation error for a sequence of NS-EW-EW station keeping maneuvers. PCRB corresponds to the standard deviation $\sigma _{\mathbf{x}} = \text{Var}\left[ ||\mathbf{x}_i^{ECI}-\hat{\mathbf{x}}^{ECI}||\right]$ where $\mathbf{x}_i^{MEE}\sim\mathcal{N}(\hat{\mathbf{x}}^{MEE},\mathcal{J}_k)$.}
    \label{fig:example_seq}
\end{figure}

The latter comes at the expense of a poor state estimation performance, as indicated by the sequential estimation error for the maneuver sequence shown in Fig. \ref{fig:example_seq}. Therein, MHE, MHE II and even the (uninformed) SHF present difficulties at recovering the orbit after a maneuver. It is not only that the state estimation error at the first post-maneuver observation is relatively high (of the order of 10 km and 1 m/s), but also two to three additional observations are usually required in order to keep the error within the variance bounds indicated by the PCRB. The proposed method, SHF II, still presents some estimation errors greater than the PCRB but is capable of rapidly converging to the true orbit as observations arrive. This faster convergence is attributed to the lower magnitude of the initial post-maneuver state error, in the order of 1 km and 0.5 m/s.

\begin{figure}[!htb]
    \centering
    \includegraphics[trim = 15 10 20 50, clip, width = .7\textwidth]{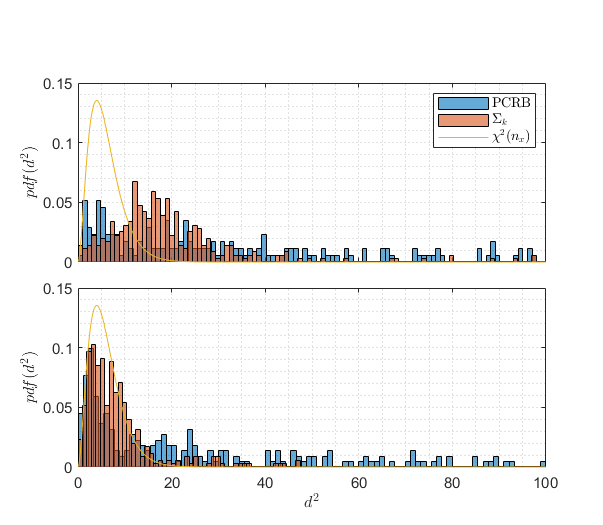}
    \caption{Statistical distribution of $d^2$ for the moving horizon estimator considering the true maneuver sequence (top), and the proposed stochastic hybrid filter with heuristics (bottom). PCRB stands for the co-variance bound defined in Eq. \eqref{eq:fim} and $\Sigma_k$ is the estimated co-variance of the state distribution.}
    \label{fig:d2_dist}
\end{figure}

\begin{table}[!htb]
    \centering
    \begin{tabular}{c|c|c|c}
        Maneuvers (Burns) & Correct & Delayed & False  \\ \hline
        MHE  &  16 (16) & 4 (4) & 0 (0) \\ \hline
        SHF II     & 20 (21)  & 0 (0) & 2 (2)  
    \end{tabular}
    \caption{Maneuver detection performance for the moving horizon estimator and the stochastic hybrid filter. The total number of detectable maneuvers is 20, whereas the number of detectable burns is 29 since EWSK require two burns and some observations are obtained between them.}
    \label{tab:mandet}
\end{table}

\begin{figure*}[!htb]
    \centering
    \hspace*{-1.2cm}
    \includegraphics[trim = 80 10 80 30, clip,width = 1.15\linewidth]{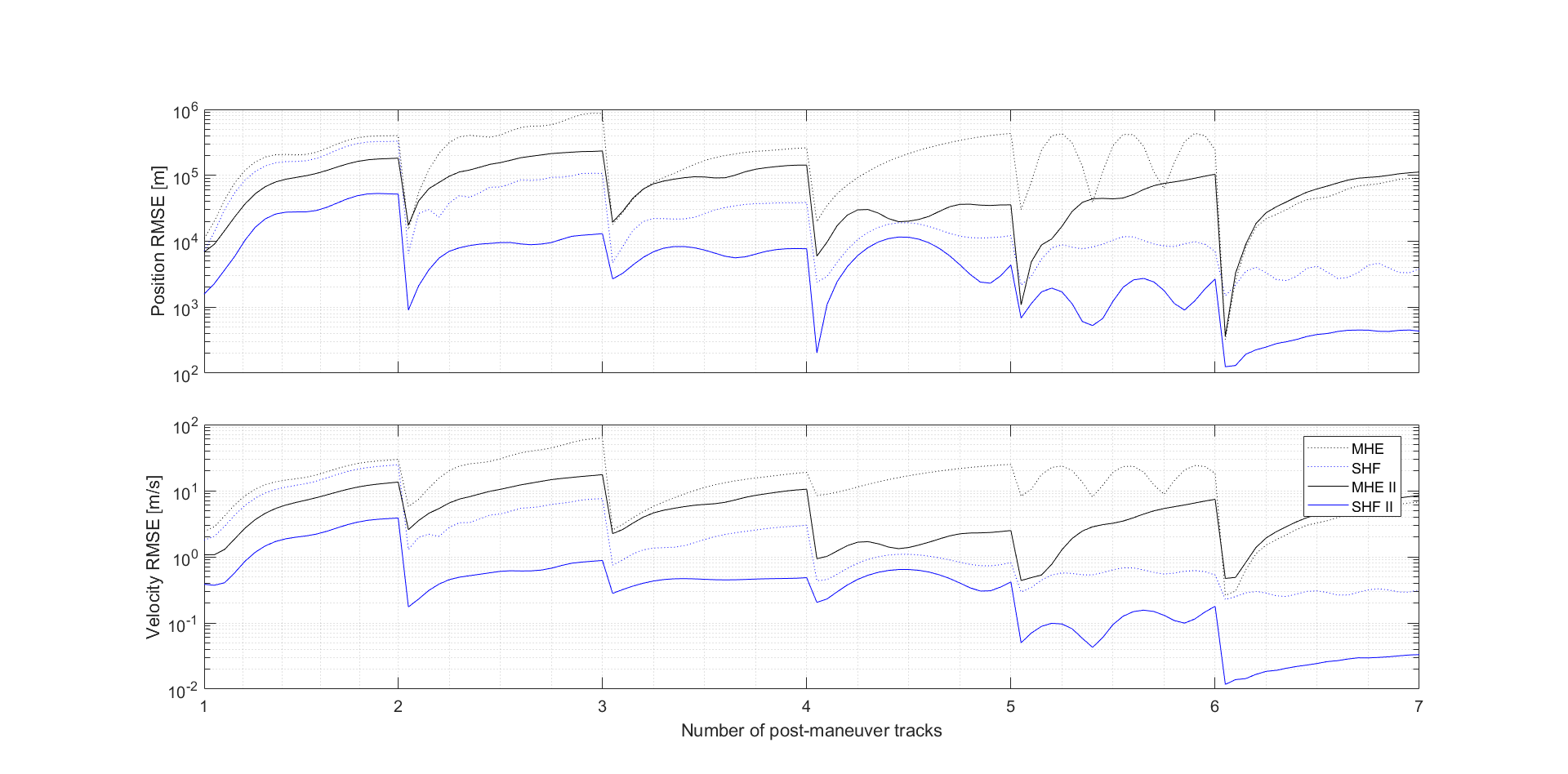}
    \caption{Online position and velocity Root Mean Square Error as a function of elapsed tracks after each maneuver. The true (and not detected) maneuver sequence is used as a reference. Note the results are averaged over a total of 20 maneuvers, yet the cardinality decreases with higher number of elapsed tracks.}
    \label{fig:rmsetracks}
\end{figure*}

\begin{table}[!htb]
    \centering
    \begin{tabular}{c|c|c|c|c}
        Method & MHE & MHE II & SHF & SHF II  \\ \hline
        Comp. Time [hrs.]  &  2.7 & 3.5 & 8.7 & 10.7 \\ 
    \end{tabular}
    \caption{Runtime comparison for the different methods. These values are obtained for a MATLAB implementation on an Intel Core i7-8750H laptop CPU.}
    \label{tab:ctime}
\end{table}

Fig. \ref{fig:d2_dist} depicts the distribution of the $d^2$ distance as defined in Eq. \eqref{eq:d2_dist} for MHE II and SHF II implementations. The conclusions that can be extracted from the pdf of $d^2$ are twofold: whether the filter is optimal in a statistical sense and how consistent is the estimated uncertainty with respect to the estimation error. The former is dictated by the PCRB, and corresponds to the blue histogram in the figure. In both cases, the distribution appears to be positively skewed with respect to the target pdf $\chi^2(n_x)$ and so either filter implementation is deemed sub-optimal. Nonetheless, SHF II presents a region of high density near the peak of the theoretical distribution, indicating a higher level of accuracy. Filter consistency is indicated by the orange bars, and can be inferred from the $d^2$ distance pdf when the estimated filter co-variance $\Sigma_k$ is used. The moving horizon estimator presents a higher skewness, so that the co-variance estimate is even lower than the PCRB. Thereafter, MHE II provides optimistic estimates as information regarding maneuvers is not included in the estimation process: the filter determines the optimal state based on minimizing the observation residuals and the co-variance is determined based on such residuals. The proposed filter, on the contrary, produces consistent state estimates since its estimated co-variance is in line with the estimation error. 

Regarding the expected estimation performance of the different filters, Fig. \ref{fig:rmsetracks} shows the \ac{RMSE} as a function of the number of tracks elapsed after every maneuver. Note the estimation error for early post-maneuver tracks is of the order of 10 km in some cases, potentially leading to a wrong track correlation and suggesting an extension of the method to multiple maneuvering targets. The scenario is similar to what was already inferred from Fig. \ref{fig:example_seq}: MHE and MHE II show a higher initial error and slower convergence than the proposed method. A possible explanation may be the presence of multiple local minima; or even a global minimum that does not necessarily correspond to the true final orbit. As the number of post-maneuver tracks increases, these local minima collapse to the solution. The proposed filter partially mitigates this problem by 1) considering multiple hypotheses, and 2) introducing prior information to explore solutions that are neither global nor local minima. The former, as shown for SHF, seems to provide a more accurate state characterization from two to five post-maneuver tracks, despite the initial error being similar to that of both MHE implementations. Including heuristics has shown to be beneficial in approximating post maneuver orbits as regions that show no relevant features a priori, are explored as indicated by the patterns of life of the target object. 

\begin{figure*}[!htb]
    \centering
    \hspace*{-2.3cm}
    \includegraphics[trim = 15 10 15 10, clip, width = 1.26\linewidth]{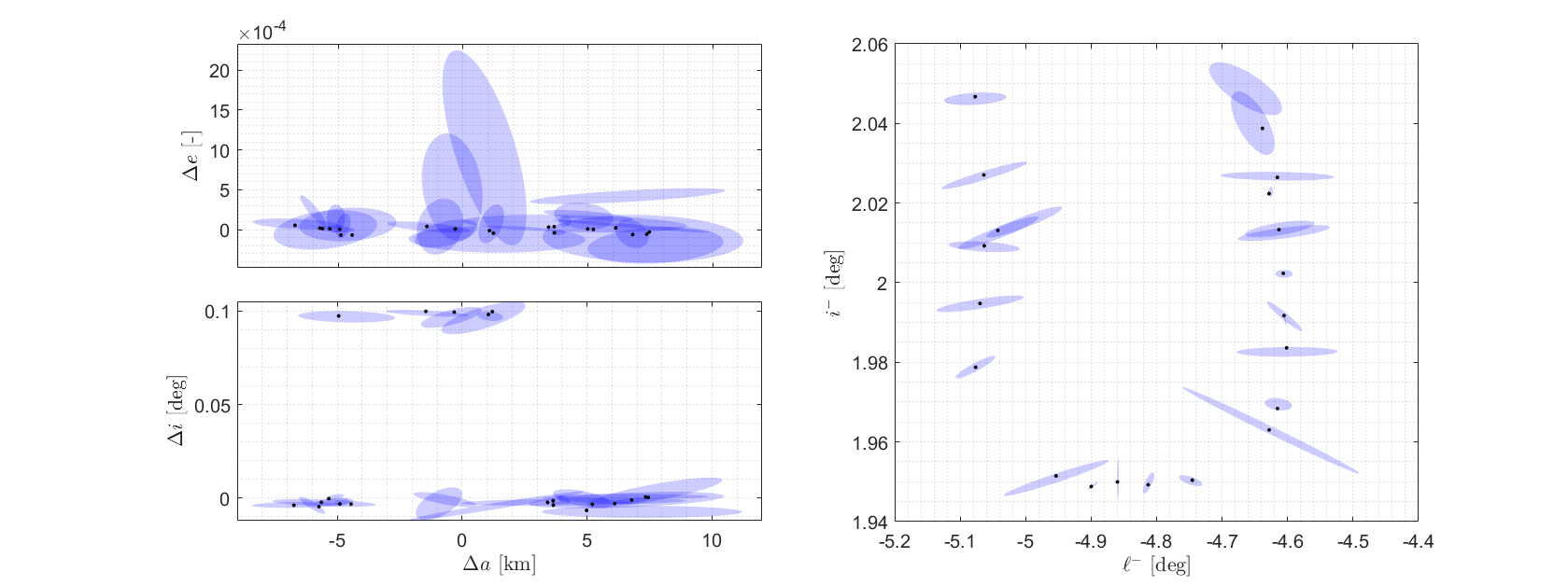}
    \caption{Maneuver characterization in terms of relative change in orbital elements (left) and pre-maneuver inclination and mean longitude (right). Blue ellipses are 3-$\sigma$ co-variance contours for the characterized maneuvers, while black dots correspond to true maneuvers.}
    \label{fig:charac_mans}
\end{figure*}

These patterns of life are summarized in Fig. \ref{fig:charac_mans}. Therein, the actual (detectable) maneuvers performed by the object are compared to the maneuvers identified and characterized by the proposed method. It can be seen that every detectable maneuver is placed within a co-variance (blue) ellipse, which coincides with the number of correctly detected maneuvers reported in Table \ref{tab:mandet}. Note that not all maneuvers are characterized with similar confidence levels, as indicated by the size of the 3-$\sigma$ co-variance bounds. In particular, there are two NSSK characterizations that show a significant standard deviation in the eccentricity when compared to the rest. One of these is directly related with a false maneuver detection, in particular the one on the top-right corner of the $\ell^-$ vs $i^-$ plot shown in Fig. \ref{fig:charac_mans}. This behavior is attributed to the lower control cost related to variations in eccentricity: orbits with higher eccentricity and lower inclination are compliant with the same tracks than higher inclination, more circular ones (if not observed at the antinodes) under certain observability conditions. Regarding EWSK maneuvers, the characterized changes in eccentricity seem to be consistent across all maneuvers, but that is not the case for the semi-major axis. The estimated $\Delta a$ values, especially those $\Delta a>0$ present a significant standard deviation, of the order of 1.3 km. Due to the positioning of the optical telescopes and the special characteristics of the GEO region, information regarding the semi-major axis is mostly conveyed in the time stamp of the observations. It is not until there is a sufficient temporal separation between tracks that the uncertainty in semi-major axis, and also eccentricity, can be reduced to acceptable levels. Accordingly, with the aim of improving the characterization of detected maneuvers, it is recommended to further reduce the uncertainty of the state estimates through a careful smoothing recursion.

In terms of computational cost, the proposed method incurs in a threefold increase with respect to the more efficient moving horizon estimator. This is partially due to the use of a regularized particle filter, which requires propagating a considerable number of state realizations $\sim 10^4$. A trade-off between uncertainty estimation accuracy and computational time may drive the use of, e.g. Gaussian Mixture Filters as in \cite{li2016gaussian}. Nonetheless, the intention of the current work is to provide a baseline for the statistical characterization of the state of a maneuvering object and so mode clustering methods as those required in Gaussian Mixture Filtering are dropped in favour of a more general statistical approximation, still at a higher computational cost.

%% file: sections/6_summary_and_conclusions.tex
\section{Summary and conclusions}\label{sec:conclusions}

A novel approach for the maneuver detection and tracking of space objects has been presented, relying on a stochastic hybrid systems formulation. Due to the scarcity of data inherent to optical space survey scenarios, estimation of the control input in the maneuvering mode is overseen in favour of a post-maneuver state estimation. The definition of an admissible control region based on a novel and efficient control distance metric has proved to be helpful in characterizing the set of feasible maneuvers performed by a target object between two subsequent tracks. Maneuver hypotheses are elaborated based on the control cost and the maneuver history, hence resulting in a combined optimal-heuristic approach. State space filtering is solved via a Regularized Particle Filter implementation, enhanced by a Markov Chain Monte Carlo characterization of post-maneuver state hypotheses.

Results are obtained for synthetic data, and comparisons are drawn against a moving horizon estimator. The proposed framework shows promising performances in post-maneuver tracking and maneuver detection accuracy, being suitable for online maneuver detection and data association purposes. The latter seem to be partially aided by the combined use of optimal control and maneuver heuristics, which helps in 1) identifying non-observable maneuvers, and 2) limiting the uncertainty in the presence of high control cost orbit updates. Moreover, success in the automation of maneuver detection and post-maneuver state estimation yields a proper characterization of the patterns of life, ultimately resulting in an increased predictability of the state of the population of active space objects.

The test scenario is representative in an operational context in terms of measurement uncertainty and environmental modeling, and so it is expected to perform similarly in an operational environment. A trade-off between computational cost and estimation accuracy may drive the use of a less demanding state space filtering technique, e.g. Gaussian Mixture Filtering, yet a careful implementation of the proposed method can enable online tracking of several targets. Certain modifications are required to adapt the methodology to multiple maneuvering target tracking, which could be based on a combination of state of the art data association methods and the proposed admissible control region.
%Within this paper, the authors propose a methodology to solve the tracking and data association problem for a single maneuvering target. This methodology, based on a stochastic hybrid systems formulation and a combined optimal-heuristic characterization of the maneuvering mode, has been successfully applied to the problem of maneuver detection and tracking of an active space object in an optical survey scenario. To determine the feasibility of the proposed method, we provide an implementation with certain approximations and assumptions.
%Future lines of research for the specific application may comprise the use of a less computationally demanding state space filtering, e.g. based on Gaussian Mixture Filtering.
%Additionally, there is room for improvement in the approach followed to numerically approximate the admissible control region, where more complex polynomial approximations, e.g. Polynomial Chaos Expansion, may result in a better characterization of the outer boundaries.
A thorough study on the different strategies that can be used to generate and apply heuristics may improve the maneuver detection and post-maneuver state estimation performances, especially if they are set according to known operator decisions or are tailored to specific procedures, e.g. limiting the accessible space to assigned orbital slots.
Moreover, under a proper characterization of the different maneuvering modes, the bi-modality of the system can be dropped in favor of a multi-modal approach, ensuring traceability in the detected maneuver type.
%Another topic of interest is the extension of the method to multiple maneuvering targets, in which the maneuver detection and complete data association (target identification) problems are jointly considered.
%Finally, the applicability of the proposal needs to be tested against real measurement data in order to validate the various approximations and assumptions, as well as the hypotheses made in the modeling of the maneuvering nature of space objects.

% contribution claim:
% GAPS:
% 1) online maneuver detection to support data association -> this goes with the increased accuracy part, if we are able to remain within reasonable error estimation bounds after a maneuver then we can say that we perform online tracking and maneuver detection (in fact we do it quite well)
% 2) combined use of optimal control and heuristics in MH framework
% 3) proper uncertainty characterization when maneuver is detected
%      This is quite relevant since after a maneuver there is a small number of observations and there is a huge uncertainty along the range and range rate...
% 4) feedback loop for characterized maneuvers: whenever a maneuver is properly characterized (at smoothing passes) the we include it in the heuristics module so we can use it later to predict future maneuvers and to characterize the post-maneuver uncertainty conditioned on historical data!!